\documentclass[aps,prd,floatfix,superscriptaddress,preprintnumbers,nofootinbib,12pt]{revtex4-1}
\usepackage{epsfig,latexsym,cancel,amssymb,amsmath,verbatim,mathrsfs}
\usepackage{xcolor}
\allowdisplaybreaks[4]
\usepackage{subfigure}
\usepackage{graphicx}
\usepackage{bm}
\usepackage{color}
\usepackage{extarrows}
\usepackage{physics}
\usepackage{booktabs} 
\usepackage{makecell}

\usepackage{multirow}

\definecolor{ccblue}{rgb}{0.0,0.4,0.8} 
\usepackage[]{hyperref}
\hypersetup{  colorlinks=true,
	linkcolor=ccblue,
	urlcolor=ccblue,
	citecolor=ccblue}

\usepackage{tikz}
\usetikzlibrary{arrows,shapes}
\usetikzlibrary{trees}
\usetikzlibrary{matrix,arrows} 				% For commutative diagram
\usetikzlibrary{positioning}				% For "above of=" commands
\usetikzlibrary{calc,through}				% For coordinates
\usetikzlibrary{decorations.pathreplacing}  % For curly braces
\usepackage{pgffor}							% For repeating patterns
\usetikzlibrary{decorations.pathmorphing}	% For Feynman Diagrams
\usetikzlibrary{decorations.markings}
\tikzset{
	vector/.style={decorate, decoration={snake}, draw},
	provector/.style={decorate, decoration={snake,amplitude=2.5pt}, draw},
	antivector/.style={decorate, decoration={snake,amplitude=-2.5pt}, draw},
	fermion/.style={draw=black, postaction={decorate},
		decoration={markings,mark=at position .55 with {\arrow[draw=black]{>}}}},
	fermionbar/.style={draw=black, postaction={decorate},
		decoration={markings,mark=at position .55 with {\arrow[draw=black]{<}}}},
	fermionnoarrow/.style={draw=black},
	gluon/.style={decorate, draw=black,
		decoration={coil,amplitude=4pt, segment length=5pt}},
	scalar/.style={dashed,draw=black, postaction={decorate},
		decoration={markings,mark=at position .55 with {\arrow[draw=black]{>}}}},
	scalarbar/.style={dashed,draw=black, postaction={decorate},
		decoration={markings,mark=at position .55 with {\arrow[draw=black]{<}}}},
	scalarnoarrow/.style={dashed,draw=black},
	electron/.style={draw=black, postaction={decorate},
		decoration={markings,mark=at position .55 with {\arrow[draw=black]{>}}}},
	bigvector/.style={decorate, decoration={snake,amplitude=4pt}, draw},
}
\tikzstyle{block} = [draw, rectangle, 
minimum height=3em, minimum width=6em]
\begin{document}
	\title{Two-Component Dark Matter with an SU(2) Dark Sector}
	\author{Shao-Long Chen}
	\email[E-mail: ]{chensl@ccnu.edu.cn}
	\author{Wen-wen Jiang}
	\email[E-Mail: ]{wwjiang@mails.ccnu.edu.cn}
	\affiliation{Key Laboratory of Quark and Lepton Physics (MoE) and Institute of Particle Physics, Central China Normal University, Wuhan 430079, China}

	\begin{abstract}
		We propose an extension to the standard model incorporating a dark sector with a non-Abelian SU(2) gauge symmetry. The model yields stable dark matter candidates, protected by a residual $Z_3$ symmetry arising after the spontaneous symmetry breaking. The dark sector interacts with the SM via a Higgs portal, facilitated from mixing between the SM Higgs doublet and a dark scalar singlet. The model features two distinct DM components. We analyze theoretical and experimental constraints, including perturbativity, unitarity, vacuum stability, dark matter relic density, direct detection, indirect detection, Higgs invisible decays, dark radiation, and ellipticity. Our findings identify viable parameter spaces that satisfy these constraints, as exemplified by two benchmark points.
	\end{abstract}

	\maketitle
	\section{Introduction}
	In the past few decades, considerable efforts have been devoted to explore the nature of dark matter (DM), including its species, mass, stability and the interactions with the standard model (SM) particles. Advances in experimental methods and technology have enabled numerous detection experiments, yet no conclusive signal has been observed, motivating the exploration of unconventional DM models.
	
	Additionally, providing the stability of DM is an important and intriguing way to understand the physics beyond the SM.  In this context, extending the SM with a dark sector endowed with an additional symmetry provides a natural framework. Drawing inspiration from stable particles within the SM itself, we consider whether DM stability can similarly arise from the remnant of a spontaneously broken dark gauge symmetry.
	
	Hidden sectors are widely applied in the physics beyond the SM. A simple extension is to consider extending the SM with a dark $ U(1)_{D} $ sector, which may or may not be spontaneously broken into a smaller symmetry~\cite{Pospelov:2007mp, Chu:2011be, Choi:2021yps,Das:2022oyx,Ko:2016ala}. Alternatively, one may consider models with other symmetries that endow dark matter with stability~\cite{Choi:2015bya,Belanger:2020hyh,Bhattacharya:2018cgx,Arcadi:2016kmk}.  Here we explore a scenario that the SM is extended with an extra dark sector possessing a non-Abelian $ SU(2)_{D} $ symmetry. The dark gauge symmetry is spontaneously broken, resulting in a residual $ Z_{3} $ symmetry that ensures the stability of the DM. The dark sector contains a triplet $\varphi$, a doublet $\chi$, and a singlet $\phi$. The singlet mixes with the SM Higgs doublet, playing the role of a mediator between the two sectors. In this model, the dark charged particles $\rho_1^{(*)}$ and the dark gauge bosons $X^{\pm}$ serve as DM candidates. Various models with a hidden $SU(2)_{D}$ have been studied in~\cite{Baek:2013dwa,Borah:2022dbw, Frigerio:2022kyu, Otsuka:2022zdy, Borah:2022phw, Baouche:2021wwa,Ko:2016yfb, Hambye:2008bq, Chen:2009ab, Diaz-Cruz:2010czr, Chowdhury:2021tnm, Ko:2020qlt, Choi:2017zww, Choi:2019zeb,Coleppa:2023vfh}. Depending on the mass difference between $X^{\pm}$ and $\rho_{1}^{(*)}$, we obtain a two-component DM scenario, which exhibits a significally different parameter space compared to the single-component thermal relic case.
	
	The paper is structured as follows. In Section II, we introduce the model and identify the DM candidates within the multi-component framework. In section III, we explore the phenomenological implications and constraints: we solve the Boltzmann equations to determine the DM relic density, provide two benchmark points, and examine theoretical constraints (such as perturbativity and unitarity), as well as the experimental constraints from DM direct detection and indirect detection. We then present our results, incorporating all the aforementioned constraints. Finally, we conclude in Section IV.
	
	\section{The Model}
	The proposed model can be seen as an extension of the SM with a gauged dark sector on $SU(2)_{D}$. The dark sector consists of one scalar triplet ($\varphi$), one scalar doublet ($\chi$) and one scalar singlet ($\phi$), which mixes with the SM Higgs and acts as a portal between the two sectors. With a non-zero vacuum expectation value(VEV) for the triplet, the theory retains a residual $Z_3$ symmetry that ensures the stability of the DM particles. The multiplets are defined as: 
	\begin{equation}
		\chi = \left(\begin{array}{c}
			\chi_{1}^{1/2}\\
			\chi_{2}^{-1/2}
		\end{array}\right)\,,\quad	
		\Phi = \frac{1}{\sqrt{2}}\left(\begin{array}{rr}
			\phi^{0}&\phi^{+} \\
			\phi^{-}& -\phi^{0}
		\end{array}\right)\,,
	\end{equation}
	where the superscripts denote the dark charge. The triplet form originates from $\Phi\equiv\varphi^{i}\sigma_{i}/\sqrt{2}$ with $\sigma_{i}$ the Pauli matrices, $\phi^0=\varphi^3$, $\phi^{\pm}=\varphi^1\mp i\varphi^2$. The Lagrangian is given by:
	\begin{align}
		\mathcal{L} =& \mathcal{L}_{\text{SM}}+
		\mathcal{L}_{d-c}+\mathcal{L}_{\phi}+\mathcal{L}_{p}\,,\notag\\
		\mathcal{L}_{d-c}=& %kinetic term 2106.10451
		-\frac{1}{4}(X^{a}_{\mu\nu}X^{a\mu\nu})+\operatorname{Tr}\left[(D_{\mu}\Phi)^{\dagger}(D^{\mu}\Phi)\right]+(D_{\mu}\chi)^{\dagger}(D^{\mu}\chi)\notag\\
		&+\frac{\mu_{\Phi}^2}{2}\operatorname{Tr}\left[\Phi^{\dagger}\Phi\right]
		-\frac{\lambda_{\Phi}}{4}(\operatorname{Tr}\left[\Phi^{\dagger}\Phi\right])^{2}
		+\frac{\mu_{\chi}^2}{2}\chi^{\dagger}\chi-\frac{\lambda_{\chi}}{4}(\chi^{\dagger}\chi)^{2}-\lambda_{\Phi\chi}\operatorname{Tr}\left[(\Phi^{\dagger}\Phi)\right](\chi^{\dagger}\chi)\notag\\
		&-(\kappa_{1}\tilde{\chi}^{\dagger}\Phi\chi+h.c.)+\kappa_{2}\chi^{\dagger}\Phi\chi
		\,,	\label{Lagrangian}\\
		\mathcal{L}_{\phi}=&(\partial_{\mu}\phi)^{\dagger}(\partial^{\mu}\phi)+\frac{\mu_{\phi}^2}{2}\phi^2-\frac{\kappa_{\phi}}{3}\phi^3-\frac{\lambda_{\phi}}{4}\phi^4-\kappa_{\phi\Phi}\phi\operatorname{Tr}\left[\Phi^{\dagger}\Phi\right]-\lambda_{\phi\Phi}\phi^{2}\operatorname{Tr}\left[\Phi^{\dagger}\Phi\right]\notag\\
		&-\kappa_{\phi\chi}\phi(\chi^{\dagger}\chi)-\lambda_{\phi\chi}\phi^{2}(\chi^{\dagger}\chi)-(\lambda_{s1}\phi\tilde{\chi}^{\dagger}\Phi\chi+h.c.)+\lambda_{s2}\phi\chi^{\dagger}\Phi\chi\,,\notag\\
		\mathcal{L}_{p}=&-\kappa_{\phi H}\phi(H^{\dagger}H)-\lambda_{\phi H}\phi^{2}(H^{\dagger}H)\notag
		\,.
	\end{align}
	Here $ X^{a}_{\mu\nu}=\partial_{\mu}X_{\nu}^{a}-\partial_{\nu}X_{\mu}^{a}+g_{D}f^{abc}X_{\mu}^{b}X_{\nu}^{c} $ is the field strength tensor of $SU(2)_D$ gauge field,  $ D_{\mu}= \partial_{\mu}-ig_{D} T^{a}X^{a}_{\mu}$ is the covariant derivative with $T^{a}=\sigma_{a}/2$ as the generator of $SU(2)_D$. We define $X_{\mu}^{\pm}=(X^{1}_{\mu}\mp iX^{2}_{\mu})/\sqrt{2}$
	and $ \tilde{\chi}=i\sigma_2\chi^{*} $. The last two terms in $\mathcal{L}_{\phi}$, $\phi \tilde{\chi}^{\dagger}\Phi\chi+h.c.$ and $\phi\chi^{\dagger}\Phi\chi$ induce notable processes where two DM particles convert into one DM particle, which we refer to as semi-annihilation processes.  The particle content and their quantum numbers are listed in Table~\ref{1}.
	\begin{table}[ht]
		\centering
		\begin{tabular}{|c|c|c|c|c|c|}
			\hline &~~ $X^{i}_{\mu}$~~&~~ $\Phi$~~ & ~~$\chi$~~ & ~~$\phi$~~& ~~$H$~~ \\
			\hline$S U(2)_{L}$ & 1 & 1 & 1 & 1 & 2 \\
			\hline$U(1)_Y$ & 0 & 0 & 0 & 0 & 1 \\
			\hline$S U(2)_{D}$ & 3 & 3 & 2 & 1 & 1 \\
			\hline
		\end{tabular}
		\caption{Particle content of the SM extended with a dark $ SU(2)_{D} $ sector. $X^{i}_{\mu}$ compose the gauge bosons of the dark sector. The scalar part consists of a triplet $ \Phi $, a doublet $ \chi $, and a real scalar $ \phi $, all singlets under the SM gauge group. $H$ is the SM Higgs doublet.}
		\label{1}
	\end{table}
	
	We choose the VEV of the doublet $ \langle\chi\rangle=0$. The triplet $ \Phi $ develops a non-zero VEV as $ \phi^{0}=v_{D}+\rho_{0} $. In this case, we get $\phi^{\pm}$ as the Nambu-Goldstone bosons  absorbed by the gauge fields $X^{\pm}$. A residual $Z_3$ symmetry remains, under which the fields transform as: $\chi_{1}^{1/2}\sim\omega^2$, $\chi_{2}^{-1/2}\sim\omega$, $\phi^0\sim1$, $\phi^+\sim\omega$ and $\phi^-\sim\omega^2$, with $\omega=e^{\pm i2\pi/3}$. 
	
	Non-diagonal mass terms in the Lagrangian indicate the mixing among $h, \phi$ and $\rho_{0}$. The squared-mass matrix for $(h, \phi, \rho_{0})$ is given by
	\begin{equation}
		-\mathcal{L} \supset \frac{1}{2}\left(\begin{array}{c}
			h \\
			\phi \\
			\rho_0
		\end{array}\right)^T\left(\begin{array}{ccc}
			M_{h}^{2} & \kappa_{\phi H}v&0 \\
			\kappa_{\phi H}v &M_{\phi}^{2}&2\kappa_{\phi\Phi}v_{D} \\
			0 & 2\kappa_{\phi\Phi}v_{D} & M_{\rho_{0}}^{2}
		\end{array}\right)\left(\begin{array}{l}
			h \\
			\phi \\
			\rho_0
		\end{array}\right)\,,
	\end{equation}
	where $M_{h}^{2}=2 \lambda_H v^2$,  $M_{\phi}^{2}=-\mu_{\phi}^{2}+\lambda_{\phi H}v^{2}+2\lambda_{\phi\Phi}v_{D}^{2}$, $M_{\rho_0}^{2}=2 \lambda_{\Phi} v_{D}^2$. 
	For simplicity, we consider the case with the assumption that $\kappa_{\phi\Phi}=0$. In this scenario, only $\phi$ and $h$ mix, while $\rho_{0}$ is the mass eigenstate. The mass eigenstates are given as
	\begin{equation}
		\left(\begin{array}{c}
			h_{1}\\
			h_{2}
		\end{array}\right) =\left(\begin{array}{cc}
			\cos{\alpha} & -\sin{\alpha}\\
			\sin{\alpha} & \cos{\alpha}
		\end{array}\right) \left(\begin{array}{c}
			h\\
			\phi
		\end{array}\right)\,,
	\end{equation}
	with the masses of the physical states $h_{1}$ and $h_{2}$ 
	\begin{equation}
		m_{h_{1,2}}^{2}=\frac{1}{2 }(M_{h}^{2}+M_{\phi}^{2})\mp\frac{1}{2}\sqrt{(M_{h}^{2}-M_{\phi}^{2})^{2}+4(\kappa_{\phi H}v)^{2}}\quad , \quad \tan 2\alpha=\frac{2\kappa_{\phi H} v}{M_{\phi}^{2}-M_{h}^{2}}\,.
	\end{equation}
	
	\subsection{The dark scalar sector}
	The quadratic terms for the dark multiplets $\Phi$ and $\chi$ are	\begin{equation}
		\begin{aligned}
			&\chi^{\dagger}\chi=(\chi_{1}^{1/2})^{*}\chi_{1}^{1/2}+(\chi_{2}^{-1/2})^{*}\chi_{2}^{-1/2}\,,\\
			&\operatorname{Tr}(\Phi^{\dagger}\Phi)=(\phi^0)^2+\phi^{+}\phi^{-}\,,
		\end{aligned}
	\end{equation}
	and the non-trivial interaction terms are
	\begin{equation}
		\begin{aligned}
			&\tilde{\chi}^{\dagger}\Phi\chi+h.c.=\frac{1}{\sqrt{2}}\left(\chi_{2}^{-1/2}\phi^0\chi_{1}^{1/2}+\chi_{2}^{-1/2}\phi^+\chi_{2}^{-1/2}-\chi_{1}^{1/2}\phi^-\chi_{1}^{1/2}+\chi_{1}^{1/2}\phi^0\chi_{2}^{-1/2}\right)+h.c.\,,\\
			&\chi^{\dagger}\Phi\chi=\frac{1}{\sqrt{2}}\left[(\chi_{1}^{1/2})^{*}\phi^0\chi_{1}^{1/2}+(\chi_{1}^{1/2})^{*}\phi^+\chi_{2}^{-1/2}+(\chi_{2}^{-1/2})^{*}\phi^-\chi_{1}^{1/2}-(\chi_{2}^{-1/2})^{*}\phi^0\chi_{2}^{-1/2}\right]\,.
		\end{aligned}
	\end{equation}
	For the doublet dark scalar, mixing occurs between its the two components, the mass matrix for ${\chi_{1}^{1/2},(\chi_{2}^{-1/2})^{*}}$ is 	\begin{equation}
		-\mathcal{L}\supset\left(\begin{array}{c}
			(\chi_{1}^{1/2})^{*}\\
			\chi_{2}^{-1/2}
		\end{array}\right)^{T} \left(\begin{array}{cc}
			M_{\chi_{1}}^{2}&\sqrt{2}\kappa_1 v_{D}\\
			\sqrt{2}\kappa_1 v_{D}&M_{\chi_{2}}^{2}
		\end{array}\right) \left(\begin{array}{c}
			\chi_{1}^{1/2}\\
			(\chi_{2}^{-1/2})^{*}
		\end{array}\right)\,,
	\end{equation}
	where $M_{\chi_{1,2}}^{2}=-\frac{\mu_{\chi}^{2}}{2}+\lambda_{\Phi\chi}v_{D}^{2}\mp\frac{1}{\sqrt{2}}\kappa_{2}v_{D}$. The mass eigenstates are
	\begin{equation}
		\left(\begin{array}{c}
			\rho_{1}\\
			\rho_{2}
		\end{array}\right) =\left(\begin{array}{cc}
			\cos{\theta} & -\sin{\theta}\\
			\sin{\theta} & \cos{\theta}
		\end{array}\right) \left(\begin{array}{c}
			\chi_{1}^{1/2}\\
			(\chi_{2}^{-1/2})^{*}
		\end{array}\right)\,,\end{equation}
	where the mass eigenstates $\rho_{1,2}$ carry a dark charge $1/2$. The mixing angle and mass eigenstates are:
	\begin{equation}
		\tan 2\theta=\frac{2\kappa_1}{\kappa_2} \,, \quad m_{\rho_{1,2}}^{2}=-\frac{\mu_{\chi}^{2}}{2}+\lambda_{\Phi\chi}v_{D}^{2}\mp\frac{1}{\sqrt{2}}v_{D}\sqrt{4\kappa_{1}^2+\kappa_{2}^2}\,,
	\end{equation}
	where we denote $m_{\rho_{1,2}}$ as masses of $\rho_{1,2}$ and it follows that $m_{\rho_{1}} < m_{\rho_{2}}$. We assume $-\mu_{\chi}^{2}+2\lambda_{\Phi\chi}v_{D}^{2} > \sqrt{2}v_{D}\sqrt{4\kappa_{1}^2+\kappa_{2}^2}$ to ensure positive eigenvalues.
	
	The mass of the scalar $\rho_{0}$ is obtained from the lagrangian in Eq.~(\ref{Lagrangian}),
	\begin{equation}
		m_{\rho_{0}}^{2}=2(-\frac{\mu_{\Phi}^{2}}{2}+\frac{3}{2}\lambda_{\Phi}v_{D}^{2})=2\lambda_{\Phi}v_{D}^{2}\,,
	\end{equation}
	where the second equality use the minimization condition $\mu_{\Phi}^{2}=\lambda_{\Phi}v_{D}^{2}$.

	\subsection{The dark gauge sector}
	The $SU(2)_{D}$ dark gauge symmetry is spontaneously broken to a residual $Z_{3}$ symmetry. Two gauge fields acquire masses by absorbing the Nambu-Goldstone bosons, while the third remains massless. The massive dark gauge bosons $X^{\pm}$ carry $Z_{3}$ charge and thus cannot decay into SM particles, which makes them stable dark matter candidates. The gauge fields couple to the scalar fields through the covariant derivative terms
	\begin{equation}
		\mathcal{L}_{\text{int}}^{g-s} =\operatorname{Tr}\left[(D_{\mu}\Phi)^{\dagger}(D^{\mu}\Phi)\right]+(D_{\mu}\chi)^{\dagger}(D^{\mu}\chi)\,.
	\end{equation}
	The covariant derivatives of the scalar fields are 
	\begin{equation}
		\begin{aligned}
			&D_{\mu}\chi =\left[\partial_\mu-ig_D
			\begin{pmatrix}
				\frac{X_{3\mu}}{2}&\frac{X^{+}_\mu}{\sqrt{2}}\\
				\frac{X^{-}_\mu}{\sqrt{2}}&-\frac{X_{3\mu}}{2}
			\end{pmatrix}\right]
			\begin{pmatrix}
				\chi_{1}^{1/2}\\
				\chi_{2}^{-1/2}
			\end{pmatrix}, \\
			&D_{\mu}\Phi =\frac{\partial_{\mu}}{\sqrt{2}}
			\begin{pmatrix}
				\phi^{0}&\phi^{+} \\
				\phi^{-}& -\phi^{0}
			\end{pmatrix}
			-\frac{ig_D}{\sqrt{2}}\left[
			\begin{pmatrix}\frac{X_{3\mu}}{2}&\frac{X^{+}_\mu}{\sqrt{2}}\\\frac{X^{-}_\mu}{\sqrt{2}}&-\frac{X_{3\mu}}{2}
			\end{pmatrix},
			\begin{pmatrix}
				\phi^{0}&\phi^{+} \\
				\phi^{-}& -\phi^{0}
			\end{pmatrix}\right]\,.
		\end{aligned}
	\end{equation}
	The dark sector gauge-scalar interactions are given by 
	\begin{align}
		\mathcal{L}_{\text{int}}^{g-s} &=(\partial^{\mu}\phi^{0})^{2}+(\partial^{\mu}\phi^{+})(\partial^{\mu}\phi^{-})\notag\\
		&+\left[-ig_{D}\partial^{\mu}\phi^{+}(\sqrt{2}X_{\mu}^{-}\phi^{0}-X_{3}X_{\mu}^{-})+h.c.\right]-\sqrt{2}ig_{D}\partial^{\mu}\phi^{0}(X_{\mu}^{+}\phi^{-}-X_{\mu}^{-}\phi^{+})\notag\\
		&-g_{D}^{2}[-2X_{\mu}^{+}X_{\mu}^{-}(\phi^{0})^{2}-X_{3}^{2}\phi^{+}\phi^{-}+\sqrt{2}X_{3}(X_{\mu}^{+}\phi^{-}+X_{\mu}^{-}\phi^{+})] \\
		&+(\partial^{\mu}\rho_{1})(\partial^{\mu}\rho_{1}^{*})+(\partial^{\mu}\rho_{2})(\partial^{\mu}\rho_{2}^{*})+\left[\frac{i}{\sqrt{2}}g_{D}X_{\mu}^{-}(\partial^{\mu}\rho_{1}\cdot\rho_{2}-\partial^{\mu}\rho_{2}\cdot\rho_{1})+h.c.\right]\notag\\
		&+\left[\frac{i}{2}g_{D}X_{3}(\partial^{\mu}\rho_{1}\cdot\rho^{*}_{1}+\partial^{\mu}\rho_{2}\cdot\rho_{2}^{*})+h.c.\right]+\frac{1}{4}g_{D}^{2}X_{3}^{2}(\rho_{1}\rho_{1}^{*}+\rho_{2}\rho_{2}^{*})\notag\\
		&+\frac{1}{2}g_{D}^{2}X_{\mu}^{+}X_{\mu}^{-}(\rho_{1}\rho_{1}^{*}+\rho_{2}\rho_{2}^{*})\,.\notag
	\end{align}
	The gauge boson $X_3$ remains massless, while the masses of the charged gauge bosons are 
	\begin{equation}	 
		m_{X^{\pm}}^{2}=2g_{D}^{2}v_{D}^{2}\,.
	\end{equation}
	
	\section{Phenomenology and Constraints}
	\subsection{Dark matter}
	After the spontaneous dark gauge symmetry breaking, the particles carrying non-zero 
	$Z_3$ charge can serve as dark matter components. In this model, the dark gauge boson $X^{\pm}$ and $\rho_{1,2}^{(*)}$ carry $Z_3$ charge. Since $\rho_1$ is lighter than $\rho_2$, there are two possible two-component DM configurations: one with $\rho_1^{(*)}$ and $X^{\pm}$, another with $\rho_1^{(*)}$ and $\rho_2^{(*)}$. We focus on the first case, where $X^{\pm}$ and $\rho_{1}^{(*)}$ are DM particles. The interaction $X^{\mp}\rho_{1}^{(*)}\rho_{2}^{(*)}$ let $\rho_2$ decay to $\rho_1^{*}X^{+}$. For this process to be kinetically allowed, we set the mass hierarchy 
	\begin{equation}
		m_{X^{\pm}}<m_{\rho_{2}}-m_{\rho_{1}}\,.
	\end{equation}
	Adopting the SM value $m_{h_{1}}=125~\rm{GeV}$, the independent parameters are: 
	\begin{equation}
		\begin{aligned}
			g_{D}\,,v_{D}\,,\sin{\theta}\,,\sin{\alpha}&\,,m_{\rho_1}\,,m_{\rho_2}\,,m_{h_2}\,,m_{\rho_0}\,,\\
			\lambda_{s1}\,,\lambda_{s2}\,,\lambda_{\chi}\,,\lambda_{\Phi\chi}\,,\kappa_{\phi}\,,\lambda_{\phi}&\,,\lambda_{\phi\Phi}\,,\kappa_{\phi\chi}\,,\lambda_{\phi\chi}\,,\lambda_{\phi H}\,.
		\end{aligned}\label{parameters}
	\end{equation}
	For brevity, we select two benchmark points (BPs), with input parameters listed in Table~\ref{benchmark points}. Two benchmark points are chosen to illustrate scenarios with different dominant dark matter components.
	\begin{table}[ht]
		\centering
		\begin{tabular}{llcc}
			%\toprule
			\Xhline{1.1pt} 
			\textbf{Parameters} & & \textbf{BP1} & \textbf{BP2} \\
			%\midrule
			\hline
			$g_D$       &   & 0.15   & 0.099 \\
			$v_D$       & [GeV] & 1000  & 1000 \\
			$\sin \alpha$ &   & 0.1    & 0.1 \\
			$\sin \theta$ &   & 0.2    & 0.2 \\
			$m_{h_1}$  & [GeV] & 125   & 125 \\
			$m_{h_2}$  & [GeV] & 200   & 200 \\
			$m_{\rho_1^{(*)}}$ & [GeV] & 200   & 300 \\
			$m_{\rho_2^{(*)}}$ & [GeV] & 3500  & 3500 \\
			$m_{\rho_0}$       & [GeV] & 1000  & 1000 \\
			%\midrule
			\hline
			$\lambda_{s1}$ & & 0.01  & 0.02 \\
			$\lambda_{s2}$ & & 0.01  & 0.01 \\
			$\lambda_{\chi}$ & & 0.1 & 0.1 \\
			$\lambda_{\Phi\chi}$ & & 0.1 & 0.1 \\
			$\kappa_\phi$ & [GeV]& 0.1 & 0.1 \\
			$\lambda_\phi$ & & 0.1 & 0.1 \\
			$\lambda_{\phi \Phi}$ & & 0.01 & 0.08 \\
			$\kappa_{\phi \chi}$ &[GeV] & 0.1 & 0.1 \\
			$\lambda_{\phi \chi}$ & & 0.18 & 0.15 \\
			$\lambda_{\phi H}$ & & 0.1 & 0.1 \\
			%\midrule
			\hline
			\textbf{Observables} & & & \\
			$m_{X^\pm}$ & [GeV] & 212 & 140 \\
			$\Omega_{\rho_{1}^{(*)}} h^{2}$ & &$1.03\times 10^{-1}$ & $8.19\times 10^{-4}$ \\
			$\Omega_{X^{\pm}} h^{2}$ & & $1.80\times 10^{-2}$  & $1.19\times 10^{-1}$ \\
			$\Omega_{\text{DM}} h^2$ & & $1.21\times 10^{-1}$ & $1.19\times 10^{-1}$ \\
			$\sigma_{\rm SI}$ & [cm$^2$] & $1.80\times 10^{-48}$ & $8.54\times 10^{-48}$ \\
			\Xhline{1.1pt}
		\end{tabular}
		\caption{The input parameters and some calculated observables are shown with two benchmark points (BPs).}
		\label{benchmark points}
	\end{table}
	\subsection{Relic density}
	We compute the evolution of the DM number density by identifying all relevant processes that modify these densities, formulating the corresponding Boltzmann equations, and numerically solving them for the relic abundance of the DM. We assume $n_{X^{+}}=n_{X^{-}}$ and $n_{\rho_{1}}=n_{\rho_{1}^{*}}$. The total relic abundance is obtained by multiplying the solution by a factor of two to account for the contributions from the antiparticles.
	
	The tree-level interactions changing the number density of $\rho_{1}$ and $X^{+}$ are 
	\begin{equation}
		\begin{aligned}
			\rho_{1}\rho_{1}^{*}&\to h_{1} h_{1}\,, h_{1} h_{2}\,,h_{2} h_{2}\,,W^{+}W^{-}\,,Z Z\,,f \bar{f}\,,X^{+}X^{-}\,, X_{3}X_{3}\,;\\
			\rho_{1}\rho_{1}&\to X^{+}X_{3}\,;\\ 
			X^{+}\rho_{1}^{*}&\to \rho_{1}X_{3}\,,\rho_{1}h_{1}\,,\rho_{1}h_{2}\,;\\
			X^{+}X^{-}&\to h_{1} h_{1}\,,h_{1} h_{2}\,, h_{2} h_{2}\,, X_{3}X_{3}\,,\rho_{1}\rho^{*}_{1}\,.
		\end{aligned}
		\label{annihilation processes}
	\end{equation}
	Processes like $X^{+}\rho_{1}^{*}\to\{\rho_{1}X_{3},~\rho_{1}h_{1},~\rho_{1}h_{2}\}$ conserve the total number density of $\rho_{1}+\rho_{1}^{*}$, so their contributions cancel in the Boltzmann equation for $\rho_1$. However, these processes do affect the number density of $X^{+}$ and must be included in the Boltzmann equation for $X^{+}$. The Boltzmann equations are	\begin{equation}
		\dv{n_{\rho_{1}}}{t}+3Hn_{\rho_1}=\sum_{i}\mathcal{C}_{\rho_1}^{i}\,,\quad\dv{n_{X^{+}}}{t}+3Hn_{X^{+}}=\sum_{i}\mathcal{C}_{X^{+}}^{i}\,,
	\end{equation}
	where $H$ is the Hubble constant. The collision terms for processes changing the number density of $\rho_{1}$ are:
	\begin{equation}
		\begin{aligned}
			&\mathcal{C}_{\rho_{1}\rho_{1}^{*}\rightarrow h_{i}h_{j}}=-\langle\sigma v_{\text{rel}}\rangle_{\rho_{1}\rho_{1}^{*}\rightarrow h_{i}h_{j}}\lbrack n_{\rho_{1}}^{2}-\frac{n_{\rho_{1}}^{\text{eq}~2}}{n^{\text{eq}}_{h_{i}}n^{\text{eq}}_{h_{j}}}n_{h_{i}}n_{h_{j}}\rbrack\,,\\
			&\mathcal{C}_{\rho_{1}\rho_{1}^{*}\rightarrow \text{SM SM}}=-\langle\sigma v_{\text{rel}}\rangle_{\rho_{1}\rho_{1}^{*}\rightarrow \text{SM SM}}\lbrack n_{\rho_{1}}^{2}-n_{\rho_{1}}^{\text{eq}~2}\rbrack\,,\\
			&\mathcal{C}_{\rho_{1}\rho_{1}^{*}\rightarrow X^{+}X^{-}}=-\langle\sigma v_{\text{rel}}\rangle_{\rho_{1}\rho_{1}^{*}\rightarrow X^{+}X^{-}}\lbrack n_{\rho_{1}}^{2}-\frac{n_{\rho_{1}}^{\text{eq}~2}}{n_{X^{+}}^{\text{eq}~2}}n_{X^{+}}^{2}\rbrack\,,\\
			&\mathcal{C}_{\rho_{1}\rho_{1}^{*}\rightarrow X_{3}X_{3}}=-\langle\sigma v_{\text{rel}}\rangle_{\rho_{1}\rho_{1}^{*}\rightarrow X_{3}X_{3}}\lbrack n_{\rho_{1}}^{2}-n_{\rho_{1}}^{\text{eq}~2}\rbrack\,,
		\end{aligned}
	\end{equation}
	\begin{equation}
		\begin{aligned}
			\mathcal{C}_{\rho_{1}\rho_{1}\rightarrow X^{+}X_{3}}=-\langle\sigma v_{\text{rel}}\rangle_{\rho_{1}\rho_{1}\rightarrow X^{+}X_{3}}\lbrack n_{\rho_{1}}^{2}-\frac{n_{\rho_{1}}^{\text{eq}~2}}{n_{X^{+}}^{\rm{eq}}}n_{X^{+}}\rbrack\notag
			\,.
		\end{aligned}
	\end{equation}
	where $i,j=1,2$, and the subscripts SM represent the SM particles in the final state of annihilation. The collision terms that change the number density of $X^{+}$ are 
	\begin{equation}
		\begin{aligned}
			\mathcal{C}_{X^{+}X^{-}\rightarrow  h_{i}h_{j}}&=-\langle\sigma v_{\text{rel}}\rangle_{X^{+}X^{-}\rightarrow  h_{i}h_{j}}\lbrack n_{X^{+}}^{2}-\frac{n_{X^{+}}^{\text{eq}~2}}{n^{\text{eq}}_{h_{i}}n^{\text{eq}}_{h_{j}}}n_{h_{i}}n_{h_{j}}\rbrack\,,\\
			\mathcal{C}_{X^{+}X^{-}\rightarrow X_{3}X_{3}}&=-\langle\sigma v_{\text{rel}}\rangle_{X^{+}X^{-}\rightarrow X_{3}X_{3}}\lbrack n_{X^{+}}^{2}-n_{X_{+}}^{\text{eq}~2}\rbrack\,,\\
			\mathcal{C}_{X^{+}X^{-}\rightarrow\rho_{1}\rho^{*}_{1}}&=-\langle\sigma v_{\text{rel}}\rangle_{X^{+}X^{-}\rightarrow\rho_{1}\rho^{*}_{1}}\lbrack n_{X^{+}}^{2}-\frac{n_{X_{+}}^{\text{eq}~2}}{n_{\rho_{1}}^{\text{eq}~2}}n_{\rho_{1}}^{2}\rbrack\,,\\
			\mathcal{C}_{X^{+}\rho_{1}^{*}\rightarrow  \rho_{1}X_{3}}&=-\langle\sigma v_{\text{rel}}\rangle_{X^{+}\rho_{1}^{*}\rightarrow  \rho_{1}X_{3}}\lbrack n_{X^{+}}n_{\rho_{1}}-n_{X_{+}}^{\rm{eq}}n_{\rho_{1}}\rbrack\,,\\
			\mathcal{C}_{X^{+}\rho_{1}^{*}\rightarrow  \rho_{1}h_{i}}&=-\langle\sigma v_{\text{rel}}\rangle_{X^{+}\rho_{1}^{*}\rightarrow  \rho_{1}h_{i}}\lbrack n_{X^{+}}n_{\rho_{1}}-\frac{n_{X_{+}}^{\rm{eq}}}{n_{h_{i}}^{\rm{eq}}}n_{\rho_{1}}n_{h_{i}}\rbrack\,.
		\end{aligned}
	\end{equation}
	%%%%%%%%%%%%%
	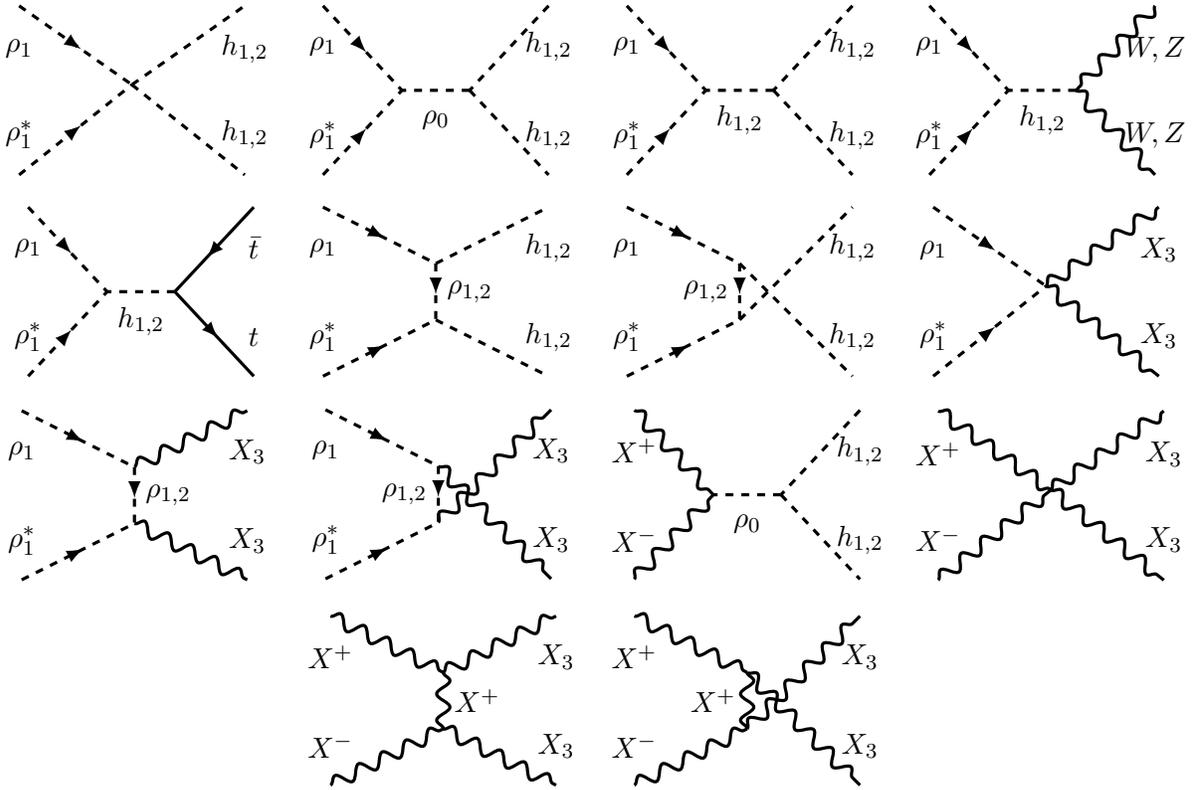
\begin{figure}[!h]
		\subfigure{
			\begin{minipage}{0.23\linewidth}
				\begin{tikzpicture}[line width=1.2 pt, scale=1.5, >=latex]
					\draw[scalar] (-1,0.5)--(0,-0.2);
					\draw[scalar] (-1,-1)--(0,-0.2);
					\draw[dashed] (0,-0.2)--(1,0.5);
					\draw[dashed] (0,-0.2)--(1,-1);
					\node at (-1,0.12) {$\rho_{1}$};
					\node at (-1,-0.64) {$\rho_{1}^{*}$};
					\node at (1,0.12) {$h_{1,2}$};
					\node at (1,-0.64) {$h_{1,2}$};
				\end{tikzpicture}
		\end{minipage}}
		%2
		\subfigure{
			\begin{minipage}{0.23\textwidth}
				\begin{tikzpicture}[line width=1.2 pt, scale=1.5, >=latex]
					\draw[scalar] (-1,0.5)--(-0.3,-0.25);
					\draw[scalar] (-1,-1)--(-0.3,-0.25);
					\draw[dashed] (0.3,-0.25)--(-0.3,-0.25);
					\draw[dashed] (1,0.5)--(0.3,-0.25);
					\draw[dashed] (0.3,-0.25)--(1,-1);
					\node at (-1,0.14) {$\rho_{1}$};
					\node at (-1,-0.65) {$\rho_{1}^{*}$};
					\node at (0,-0.51) {$\rho_{0}$};
					\node at (1,0.14) {$h_{1,2}$};
					\node at (1,-0.65) {$h_{1,2}$};
				\end{tikzpicture}
		\end{minipage}}
		%3
		\subfigure{
			\begin{minipage}{0.23\textwidth}
				\begin{tikzpicture}[line width=1.2 pt, scale=1.5, >=latex]
					\draw[scalar](-1,0.5)--(-0.3,-0.25);
					\draw[scalar] (-1,-1)--(-0.3,-0.25);
					\draw[dashed] (0.3,-0.25)--(-0.3,-0.25);
					\draw[dashed] (1,0.5)--(0.3,-0.25);
					\draw[dashed] (0.3,-0.25)--(1,-1);
					\node at (-1,0.14) {$\rho_{1}$};
					\node at (-1,-0.65) {$\rho_{1}^{*}$};
					\node at (0,-0.51) {$h_{1,2}$};
					\node at (1,0.14) {$h_{1,2}$};
					\node at (1,-0.65) {$h_{1,2}$};
				\end{tikzpicture}
		\end{minipage}}	
		%4
		\subfigure{
			\begin{minipage}{0.23\textwidth}
				\begin{tikzpicture}[line width=1.2 pt, scale=1.5, >=latex]
					\draw[scalar](-1,0.5)--(-0.3,-0.25);
					\draw[scalar] (-1,-1)--(-0.3,-0.25);
					\draw[dashed] (0.3,-0.25)--(-0.3,-0.25);
					\draw[decorate, decoration={snake}] (1,0.5)--(0.3,-0.25);
					\draw[decorate, decoration={snake}] (0.3,-0.25)--(1,-1);
					\node at (-1,0.14) {$\rho_{1}$};
					\node at (-1,-0.65) {$\rho_{1}^{*}$};
					\node at (0,-0.51) {$h_{1,2}$};
					\node at (1,0.115) {$W, Z$};
					\node at (1,-0.65) {$W, Z$};
				\end{tikzpicture}
		\end{minipage}}	
		%5
		\subfigure{
			\begin{minipage}{0.23\textwidth}
				\begin{tikzpicture}[line width=1.2 pt, scale=1.5, >=latex]
					\draw[scalar](-1,0.5)--(-0.3,-0.25);
					\draw[scalar] (-1,-1)--(-0.3,-0.25);
					\draw[dashed] (0.3,-0.25)--(-0.3,-0.25);
					\draw[fermionbar] (0.3,-0.25)--(1,0.5);
					\draw[fermion] (0.3,-0.25)--(1,-1);
					\node at (-1,0.14) {$\rho_{1}$};
					\node at (-1,-0.65) {$\rho_{1}^{*}$};
					\node at (0,-0.51) {$h_{1,2}$};
					\node at (1,0.14) {$\bar{t}$};
					\node at (1,-0.65) {$t$};
				\end{tikzpicture}
		\end{minipage}}	
		%6
		\subfigure{
			\begin{minipage}{0.23\textwidth}
				\begin{tikzpicture}[line width=1.2 pt, scale=1.5, >=latex]
					\draw[scalar] (-1,0.5)--(0,0);
					\draw[dashed] (0,0)--(1,0.5);
					\draw[scalar] (0,0)--(0,-0.5);
					\draw[scalar] (-1,-1)--(0,-0.5);
					\draw[dashed] (0,-0.5)--(1,-1);
					\node at (-1,0.14) {$\rho_{1}$};
					\node at (-1,-0.65) {$\rho_{1}^{*}$};
					\node at (0.3,-0.25) {$\rho_{1,2}$};
					\node at (1,0.14) {$h_{1,2}$};
					\node at (1,-0.65) {$h_{1,2}$};
				\end{tikzpicture}
		\end{minipage}}
		%7
		\subfigure{
			\begin{minipage}{0.23\textwidth}
				\begin{tikzpicture}[line width=1.2 pt, scale=1.5, >=latex]
					\draw[scalar] (-1,0.5)--(0,0);
					\draw[dashed] (0,0)--(1,-1);
					\draw[scalar] (0,0)--(0,-0.5);
					\draw[scalar] (-1,-1)--(0,-0.5);
					\draw[dashed] (0,-0.5)--(1,0.5);
					\node at (-1,0.14) {$\rho_{1}$};
					\node at (-1,-0.65) {$\rho_{1}^{*}$};
					\node at (-0.3,-0.25) {$\rho_{1,2}$};
					\node at (1,0.14) {$h_{1,2}$};
					\node at (1,-0.65) {$h_{1,2}$};
				\end{tikzpicture}
		\end{minipage}}
		%8
		\subfigure{
			\begin{minipage}{0.23\textwidth}
				\begin{tikzpicture}[line width=1.2 pt, scale=1.5, >=latex]
					\draw[scalar] (-1,0.5)--(0,-0.2);
					\draw[scalar] (-1,-1)--(0,-0.2);
					\draw[decorate, decoration={snake}] (0,-0.2)--(1,0.5);
					\draw[decorate, decoration={snake}] (0,-0.2)--(1,-1);
					\node at (-1,0.12) {$\rho_{1}$};
					\node at (-1,-0.64) {$\rho_{1}^{*}$};
					\node at (1,0.12) {$X_3$};
					\node at (1,-0.64) {$X_3$};
				\end{tikzpicture}
		\end{minipage}}
		%9
		\subfigure{
			\begin{minipage}{0.23\textwidth}
				\begin{tikzpicture}[line width=1.2 pt, scale=1.5, >=latex]
					\draw[scalar] (-1,0.5)--(0,0);
					\draw[decorate, decoration={snake}] (0,0)--(1,0.5);
					\draw[scalar] (0,0)--(0,-0.5);
					\draw[scalar] (-1,-1)--(0,-0.5);
					\draw[decorate, decoration={snake}] (0,-0.5)--(1,-1);
					\node at (-1,0.14) {$\rho_{1}$};
					\node at (-1,-0.65) {$\rho_{1}^{*}$};
					\node at (0.3,-0.25) {$\rho_{1,2}$};
					\node at (1,0.14) {$X_3$};
					\node at (1,-0.65) {$X_3$};
				\end{tikzpicture}
		\end{minipage}}
		%10
		\subfigure{
			\begin{minipage}{0.23\textwidth}
				\begin{tikzpicture}[line width=1.2 pt, scale=1.5, >=latex]
					\draw[scalar] (-1,0.5)--(0,0);
					\draw[decorate, decoration={snake}] (0,0)--(1,-1);
					\draw[scalar] (0,0)--(0,-0.5);
					\draw[scalar] (-1,-1)--(0,-0.5);
					\draw[decorate, decoration={snake}] (0,-0.5)--(1,0.5);
					\node at (-1,0.14) {$\rho_{1}$};
					\node at (-1,-0.65) {$\rho_{1}^{*}$};
					\node at (-0.3,-0.25) {$\rho_{1,2}$};
					\node at (1,0.14) {$X_3$};
					\node at (1,-0.65) {$X_3$};
				\end{tikzpicture}
		\end{minipage}}
		%11
		\subfigure{
			\begin{minipage}{0.23\textwidth}
				\begin{tikzpicture}[line width=1.2 pt, scale=1.5, >=latex]
					\draw[decorate, decoration={snake}] (-1,0.5)--(-0.3,-0.25);
					\draw[decorate, decoration={snake}] (-1,-1)--(-0.3,-0.25);
					\draw[dashed] (0.3,-0.25)--(-0.3,-0.25);
					\draw[dashed] (1,0.5)--(0.3,-0.25);
					\draw[dashed] (0.3,-0.25)--(1,-1);
					\node at (-1,0.14) {$X^{+}$};
					\node at (-1,-0.65) {$X^{-}$};
					\node at (0,-0.51) {$\rho_{0}$};
					\node at (1,0.14) {$h_{1,2}$};
					\node at (1,-0.65) {$h_{1,2}$};
				\end{tikzpicture}
		\end{minipage}}
		%12
		\subfigure{
			\begin{minipage}{0.23\textwidth}
				\begin{tikzpicture}[line width=1.2 pt, scale=1.5, >=latex]
					\draw[decorate, decoration={snake}] (-1,0.5)--(0,-0.2);
					\draw[decorate, decoration={snake}] (-1,-1)--(0,-0.2);
					\draw[decorate, decoration={snake}] (0,-0.2)--(1,0.5);
					\draw[decorate, decoration={snake}] (0,-0.2)--(1,-1);
					\node at (-1,0.12) {$X^{+}$};
					\node at (-1,-0.64) {$X^{-}$};
					\node at (1,0.12) {$X_3$};
					\node at (1,-0.64) {$X_3$};
				\end{tikzpicture}
		\end{minipage}}
		%13
		\subfigure{
			\begin{minipage}{0.23\textwidth}
				\begin{tikzpicture}[line width=1.2 pt, scale=1.5, >=latex]
					\draw[decorate, decoration={snake}] (-1,0.5)--(0,0);
					\draw[decorate, decoration={snake}] (0,0)--(1,0.5);
					\draw[decorate, decoration={snake}] (0,0)--(0,-0.5);
					\draw[decorate, decoration={snake}] (-1,-1)--(0,-0.5);
					\draw[decorate, decoration={snake}] (0,-0.5)--(1,-1);
					\node at (-1,0.14) {$X^{+}$};
					\node at (-1,-0.65) {$X^{-}$};
					\node at (0.3,-0.25) {$X^{+}$};
					\node at (1,0.14) {$X_3$};
					\node at (1,-0.65) {$X_3$};
				\end{tikzpicture}
		\end{minipage}}
		%14
		\subfigure{
			\begin{minipage}{0.23\textwidth}
				\begin{tikzpicture}[line width=1.2 pt, scale=1.5, >=latex]
					\draw[decorate, decoration={snake}] (-1,0.5)--(0,0);
					\draw[decorate, decoration={snake}] (0,0)--(1,-1);
					\draw[decorate, decoration={snake}] (0,0)--(0,-0.5);
					\draw[decorate, decoration={snake}] (-1,-1)--(0,-0.5);
					\draw[decorate, decoration={snake}] (0,-0.5)--(1,0.5);
					\node at (-1,0.14) {$X^{+}$};
					\node at (-1,-0.65) {$X^{-}$};
					\node at (-0.3,-0.25) {$X^{+}$};
					\node at (1,0.14) {$X_3$};
					\node at (1,-0.65) {$X_3$};
				\end{tikzpicture}
		\end{minipage}}
		\caption{The Feynman diagrams of DM annihilation processes. The relevant processes are shown in Eq.~(\ref{annihilation processes}).}
		\label{DM annihilation}
	\end{figure}
	\begin{figure}[!h]%[!t]
		%1
		\subfigure{
			\begin{minipage}{0.23\textwidth}
				\begin{tikzpicture}[line width=1.2 pt, scale=1.5, >=latex]
					\draw[scalar] (-1,0.5)--(-0.3,-0.25);
					\draw[scalar] (-1,-1)--(-0.3,-0.25);
					\draw[decorate, decoration={snake}] (0.3,-0.25)--(-0.3,-0.25);
					\draw[decorate, decoration={snake}] (1,0.5)--(0.3,-0.25);
					\draw[decorate, decoration={snake}] (0.3,-0.25)--(1,-1);
					\node at (-1,0.14) {$\rho_{1}$};
					\node at (-1,-0.65) {$\rho_{1}$};
					\node at (0,-0.51) {$X^{+}$};
					\node at (1,0.14) {$X^{+}$};
					\node at (1,-0.65) {$X_{3}$};
				\end{tikzpicture}
		\end{minipage}}
		%2
		\subfigure{
			\begin{minipage}{0.23\textwidth}
				\begin{tikzpicture}[line width=1.2 pt, scale=1.5, >=latex]
					\draw[scalar] (-1,0.5)--(0,0);
					\draw[scalar] (0,0)--(1,0.5);
					\draw[decorate, decoration={snake}]  (0,-0.5)--(0,0);
					\draw[decorate, decoration={snake}] (-1,-1)--(0,-0.5);
					\draw[decorate, decoration={snake}] (0,-0.5)--(1,-1);
					\node at (-1,0.14) {$\rho_{1}^{*}$};
					\node at (-1,-0.65) {$X^{+}$};
					\node at (0.3,-0.25) {$X^{+}$};
					\node at (1,0.14) {$\rho_{1}$};
					\node at (1,-0.65) {$X_{3}$};
				\end{tikzpicture}
		\end{minipage}}
		%3
		\subfigure{
			\begin{minipage}{0.23\textwidth}
				\begin{tikzpicture}[line width=1.2 pt, scale=1.5, >=latex]
					\draw[scalar] (-1,0.5)--(-0.3,-0.25);
					\draw[decorate, decoration={snake}] (-1,-1)--(-0.3,-0.25);
					\draw[scalar] (-0.3,-0.25)--(0.3,-0.25);
					\draw[scalar] (0.3,-0.25)--(1,0.5);
					\draw[dashed] (0.3,-0.25)--(1,-1);
					\node at (-1,0.14) {$\rho_{1}^{*}$};
					\node at (-1,-0.65) {$X^{+}$};
					\node at (0,-0.51) {$\rho_{1}$};
					\node at (1,0.14) {$\rho_{1}$};
					\node at (1,-0.65) {$h_{1,2}$};
				\end{tikzpicture}
		\end{minipage}}
		\caption{DM semi-annihilation processes in the model. }
		\label{DM semi-annihilation}
	\end{figure}
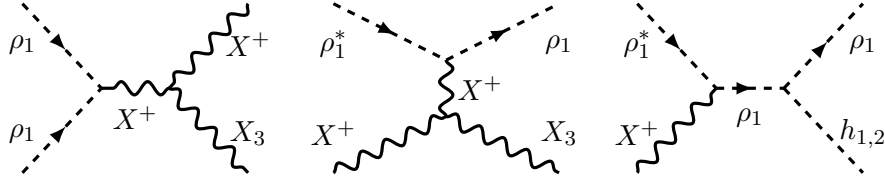
	
	\begin{figure}[!ht]%[!t]
		%1
		\subfigure{
			\begin{minipage}{0.23\textwidth}
				\begin{tikzpicture}[line width=1.2 pt, scale=1.5, >=latex]
					\draw[scalar] (-1,0.5)--(0,-0.2);
					\draw[scalar] (-1,-1)--(0,-0.2);
					\draw[decorate, decoration={snake}] (0,-0.2)--(1,0.5);
					\draw[decorate, decoration={snake}] (0,-0.2)--(1,-1);
					\node at (-1,0.12) {$\rho_{1}$};
					\node at (-1,-0.64) {$\rho_{1}^{*}$};
					\node at (1,0.12) {$X^{+}$};
					\node at (1,-0.64) {$X^{-}$};
				\end{tikzpicture}
		\end{minipage}}
		%2
		\subfigure{
			\begin{minipage}{0.23\textwidth}
				\begin{tikzpicture}[line width=1.2 pt, scale=1.5, >=latex]
					\draw[scalar] (-1,0.5)--(-0.3,-0.25);
					\draw[scalar] (-1,-1)--(-0.3,-0.25);
					\draw[dashed] (0.3,-0.25)--(-0.3,-0.25);
					\draw[decorate, decoration={snake}] (0.3,-0.25)--(1,0.5);
					\draw[decorate, decoration={snake}] (0.3,-0.25)--(1,-1);
					\node at (-1,0.14) {$\rho_{1}$};
					\node at (-1,-0.65) {$\rho_{1}^{*}$};
					\node at (0,-0.47) {$\rho_{0}$};%/X_3
					\node at (1,0.14) {$X^{+}$};
					\node at (1,-0.65) {$X^{-}$};
				\end{tikzpicture}
		\end{minipage}}
		%3
		\subfigure{
			\begin{minipage}{0.23\textwidth}
				\begin{tikzpicture}[line width=1.2 pt, scale=1.5, >=latex]
					\draw[scalar] (-1,0.5)--(-0.3,-0.25);
					\draw[scalar] (-1,-1)--(-0.3,-0.25);
					\draw[decorate, decoration={snake}] (0.3,-0.25)--(-0.3,-0.25);
					\draw[decorate, decoration={snake}] (0.3,-0.25)--(1,0.5);
					\draw[decorate, decoration={snake}] (0.3,-0.25)--(1,-1);
					\node at (-1,0.14) {$\rho_{1}$};
					\node at (-1,-0.65) {$\rho_{1}^{*}$};
					\node at (0,-0.47) {$X_3$};
					\node at (1,0.14) {$X^{+}$};
					\node at (1,-0.65) {$X^{-}$};
				\end{tikzpicture}
		\end{minipage}}
		%4
		\subfigure{
			\begin{minipage}{0.23\textwidth}
				\begin{tikzpicture}[line width=1.2 pt, scale=1.5, >=latex]
					\draw[scalar] (-1,0.5)--(0,0);
					\draw[decorate, decoration={snake}] (0,0)--(1,0.5);
					\draw[scalar] (0,-0.5)--(0,0);
					\draw[scalar] (-1,-1)--(0,-0.5);
					\draw[decorate, decoration={snake}] (0,-0.5)--(1,-1);
					\node at (-1,0.14) {$\rho_{1}$};
					\node at (-1,-0.65) {$\rho_{1}^{*}$};
					\node at (0.3,-0.25) {$\rho_{1,2}$};
					\node at (1,0.14) {$X^{+}$};
					\node at (1,-0.65) {$X^{-}$};
				\end{tikzpicture}
		\end{minipage}}	   
		\caption{The Feynman diagrams for the DM conversion processes $\rho_{1}+\rho_{1}^{*}\rightarrow X^{+}+X^{-}$, the reverse processes can be obtained by exchanging the initial and final states.}
		\label{DM conversions}
	\end{figure}
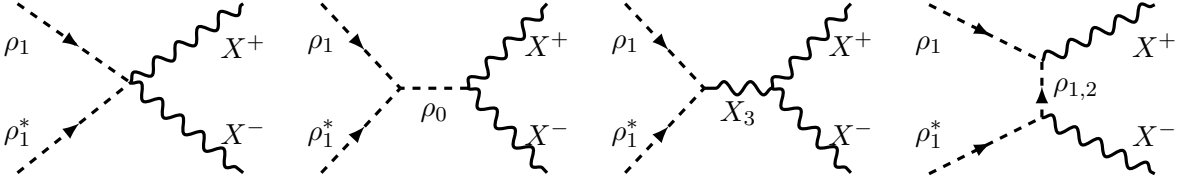
	The collision terms include a number of tree-level annihilation, semi-annihilation and conversion processes that determine the DM relics. These processes, illustrated in Fig.~\ref{DM annihilation}-\ref{DM conversions} can be categorized into contributions from direct quartic interactions and from $s,t,u$-channel diagrams. To more conveniently describe the Boltzmann equations that govern the evolution of the particle number densities for the processes shown in the figure, we replace the variables $n_{i}$ and $t$ with the yield $Y$, a rescaled time variable $x_{i}$ together with a parameter $\lambda$, defined as
	\begin{equation}
		x_{i}\equiv\frac{m_{i}}{T}\,,\quad Y_{i}\equiv\frac{n_{i}}{s}\,,\quad \lambda_{ab\rightarrow cd}\equiv\frac{s(x=1)}{H(x=1)}\langle\sigma v_{\text{rel}}\rangle_{ab\rightarrow cd}(x)\,,
	\end{equation}
	where $i=\rho_{1}, X^{+}$ denotes the DM particles, $m_{i}$ as the mass of the DM particles and $T$ as the temperature of the thermal bath. $s=(2\pi^{2}/45)g_{*s}T^3$ is the entropy density with $g_{*s}$ defined as the effective number of degrees of freedom. The detail of $\langle\sigma v_{\text{rel}}\rangle$ of different reactions are given in the Appendix. With these variables, the Boltzmann equation of $\rho_1$ and $X^{+}$ can be written as
	\begin{equation}
		\begin{aligned}
			\frac{dY_{\rho_1}}{dx_{\rho_1}}=-\frac{1}{x_{\rho_1}^{2}}\Big\lbrace
			&\lambda_{\rho_{1}\rho_{1}^{*}\rightarrow h_{i}h_{j}}\Big\lbrack Y_{\rho_{1}}^{2}
			-\frac{Y_{\rho_{1}}^{\text{eq}~2}}{Y^{\text{eq}}_{h_{i}}Y^{\text{eq}}_{h_{j}}}Y_{h_{i}}Y_{h_{j}}\Big\rbrack
			+\lambda_{\rho_{1}\rho_{1}^{*}\rightarrow SM SM}\Big\lbrack Y_{\rho_{1}}^{2}-Y_{\rho_{1}}^{\text{eq}~2}\Big\rbrack
			+\lambda_{\rho_{1}\rho_{1}^{*}\rightarrow X^{+}X^{-}}\\
			&\times\Big\lbrack Y_{\rho_{1}}^{2}-\frac{Y_{\rho_{1}}^{\text{eq}~2}}{Y_{X^{+}}^{\text{eq}~2}}Y_{X^{+}}^{2}\Big\rbrack+\lambda_{\rho_{1}\rho_{1}^{*}\rightarrow X_3 X_3}\Big\lbrack Y_{\rho_{1}}^{2}-Y_{\rho_{1}}^{\text{eq}~2}\Big\rbrack
			+\lambda_{\rho_{1}\rho_{1}\rightarrow X^{+}X_3}\\
			&\times\Big\lbrack Y_{\rho_{1}}^{2}-\frac{Y_{\rho_{1}}^{\text{eq}~2}}{Y_{X^{+}}^{\rm{eq}}}Y_{X^{+}}\Big\rbrack\Big\rbrace\,,\\
			\frac{dY_{X^{+}}}{dx_{X^{+}}}=-\frac{1}{x_{X^{+}}^{2}}\Big\lbrace
			&\lambda_{X^{+}X^{-}\rightarrow h_{i}h_{j}}\Big\lbrack Y_{X^{+}}^{2}-
			-\frac{Y_{X^{+}}^{\text{eq}~2}}{Y^{\text{eq}}_{h_{i}}Y^{\text{eq}}_{h_{j}}}Y_{h_{i}}Y_{h_{j}}\Big\rbrack+\lambda_{X^{+}X^{-}\rightarrow X_{3}X_{3}}\Big\lbrack Y_{X^{+}}^{2}-Y_{X^{+}}^{\text{eq}~2}\Big\rbrack\\
			&+\lambda_{X^{+}X^{-}\rightarrow\rho_{1}\rho^{*}_{1}}\Big\lbrack Y_{X^{+}}^{2}-\frac{Y_{X_{+}}^{\text{eq}~2}}{Y_{\rho_{1}}^{eq~2}}Y_{\rho_{1}}^{2}\Big\rbrack+\lambda_{X^{+}\rho_{1}^{*}\rightarrow\rho_{1}X_{3}}\Big\lbrack Y_{X^{+}}Y_{\rho_{1}}-Y_{X^{+}}^{\rm{eq}}Y_{\rho_{1}}\Big\rbrack\\
			&+\lambda_{X^{+}\rho_{1}^{*}\rightarrow\rho_{1}h_{i}}\Big\lbrack Y_{X^{+}}Y_{\rho_{1}}-\frac{Y_{X_{+}}^{\rm{eq}}}{Y_{h_{i}}^{\rm{eq}}}Y_{\rho_{1}}Y_{h_{i}}\Big\rbrack\Big\rbrace\,,
		\end{aligned}
	\end{equation}
	where the yield in equilibrium $Y^{\text{eq}}$ of DM particles are defined by
	\begin{equation}
		\begin{aligned}
			&Y^{\rm{eq}}_{\rho_1}\equiv\frac{n^{\rm{eq}}_{\rho_1}}{s}=\frac{45}{4\pi^4}\frac{g_{\rho_1}}{g_{*s}}x_{\rho_1}^2K_{2}\lbrack x_{\rho_1}\rbrack\,,\\
			&Y^{\rm{eq}}_{X^{+}}\equiv\frac{n^{\rm{eq}}_{X^{+}}}{s}=\frac{45}{4\pi^4}\frac{g_{\rho_1}}{g_{*s}}r^2x_{\rho_1}^2K_{2}\lbrack rx_{\rho_1}\rbrack\,,
		\end{aligned}
	\end{equation}
	where $r=m_{X^{+}}/m_{\rho_1}$ is the mass ratio of two DM particles, $K_{2}\lbrack x\rbrack$ is the modified Bessel function of the second kind of order 2.
	
	We utilize the FeynRules 2~\cite{Alloul:2013bka} and micrOMEGAs 6.0.5~\cite{Alguero:2023zol} packages to determine the annihilation cross-section and the relic density of the DMs. The sum of the relics of DM particles is compared to the observed value of PLANCK, $\Omega_{\rm{DM}}h^{2}=0.1200\pm0.0012$~\cite{Planck:2018vyg}.
	\begin{figure}[!htb]
		\begin{minipage}{0.48\linewidth}
			\centerline{\includegraphics[width=1\textwidth]{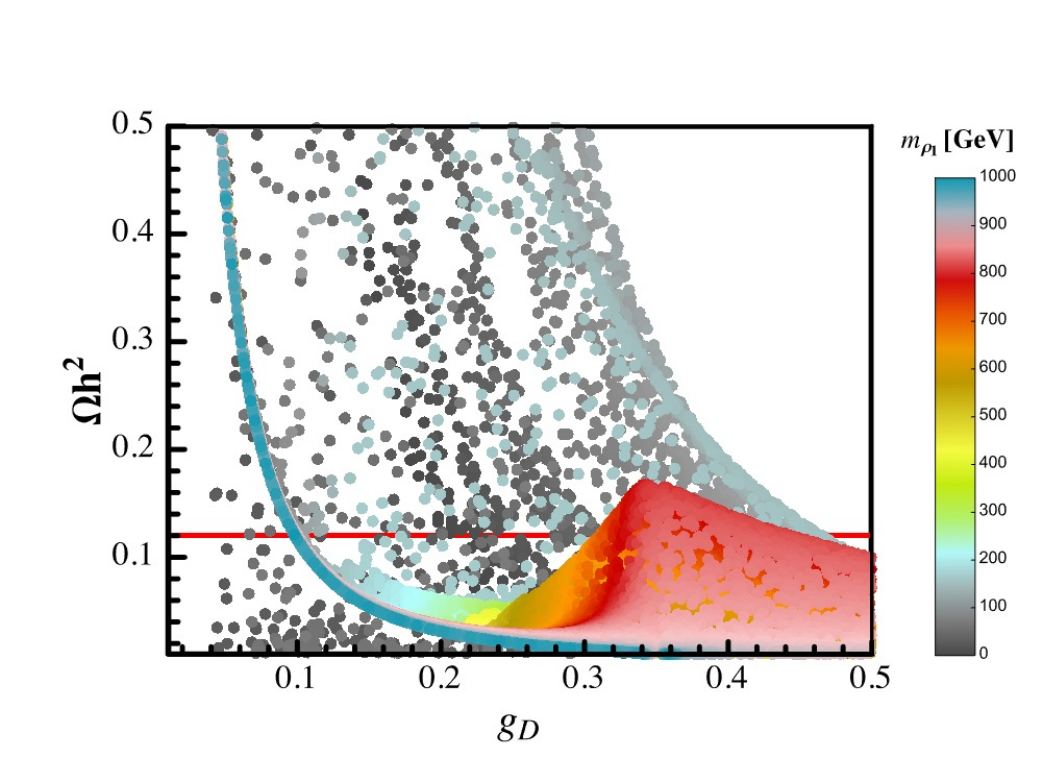}}
		\end{minipage}
		\hspace{0.01cm}
		\begin{minipage}{0.505\linewidth}   \centerline{\includegraphics[width=1.05\textwidth]{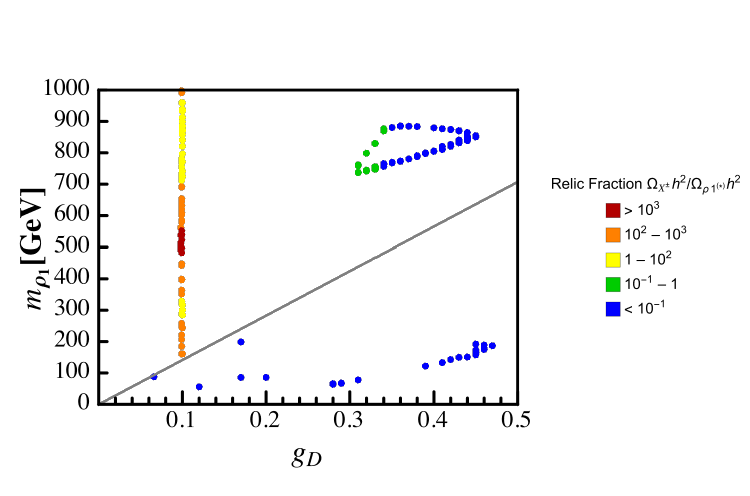}}
		\end{minipage}
		\caption{Left panel:  The DM relic density versus the value of the dark sector coupling $g_{D}$. The color bar on the right side represents the DM mass $m_{\rho_1}$. The red line refers to the relic density observed by the PLANCK. Right panel: Collections of points filtered by the observed relic density. The gray line represents the $m_{X^{\pm}}=m_{\rho_1}$ case. Different colors tracks the fraction of the DM relic density.} 
		\label{gD-relic}
	\end{figure}
	The relic density with different DM masses are shown in Fig.~\ref{gD-relic}. We scan the dark sector coupling $g_{D}\in (0,0.5)$, The value of the dark sector coupling $g_{D}$ is related with the DM mass $m_{X^{\pm}}$ by $g_{D}=m_{X^{\pm}}/(\sqrt{2}v_{D})$. Meanwhile, we scan the dark matter mass $m_{\rho_1}\in(50,1000)~\text{GeV}$, different values of the other DM mass $m_{\rho_1}$ are indicated by different colors of the points.  The settings of other parameters are consistent with BP1. For various mass combinations of the two-component DM, the corresponding relic density results are shown in the left panel. The red line corresponds to the PLANCK measurements, the points lie in the area with $g_{D}$ less than 0.47 can reproduce the right relic density. The hill on the right originates from the increasing contribution of $\rho_1$ to the total relic density. In the right panel, different colors indicate the fraction of the dark matter relic density, we can see that the points filtered by the relic density cover almost all the mass range between 100-1000 $\rm{GeV}$ when $g_{D}$ takes around 0.1. The gray line corresponds to the case where the two DM particles have equal masses. Below the gray line, where $m_{X^{\pm}}>m_{\rho_1}$, $\rho_1$ dominates the relic density, and a lighter $\rho_1$ mass can naturally yield a relic abundance around $\Omega h^{2}=0.12$. The relic density of $\rho_1$ is much larger than that of $X^{\pm}$. In contrast, above the gray line, where $m_{\rho_1}>m_{X^{\pm}}$, the dark gauge boson $X^{\pm}$ becomes the dominant DM component, and for coupling values around $g_{D}\sim 0.1$, it can account for the observed relic density, $\Omega_{X^{\pm}}$ dominates the total relic density.
	
	\subsection{Perturbativity, Unitarity and Vacuum stability}
	We consider the constraints on the model parameters from the following theoretical requirements. 
	
	\noindent \textbf{Perturbativity:} To ensure that the one-loop level quantum corrections remains smaller than the tree level contributions, all vertex couplings should be less than $4\pi$. 
	
	\noindent \textbf{Unitarity:} At high energies, the quartic couplings dominantly contribute to the amplitudes of $2\rightarrow2$ scattering~\cite{Arhrib:2000is}. Hence, the scattering amplitudes that quartic couplings $\lambda_{\Phi}$, $\lambda_{\chi}$, $\lambda_{\phi}$, $\lambda_{\Phi\chi}$, $\lambda_{\phi\Phi}$, $\lambda_{\phi\chi}$, $\lambda_{\phi H}$, $\lambda_{s1}$ and $\lambda_{s2}$ contribute to should respect the tree-level unitarity at high energies and give the constraints. Denoting the eigenvalues of the scattering matrix as $\Lambda_{i}$, the unitarity condition is governed by $\left|\Lambda_{i}\right|\leq8\pi$.
	
	\noindent \textbf{Vacuum stability:} The scalar potential must be bounded from below to guarantee a stable minimum for perturbative calculations. As for the potential in this model, the stability conditions do not constrain couplings $\lambda_{s1}$ and $\lambda_{s2}\,.$ The symmetric matrix of the other quartic couplings is:
	\begin{equation}
		M=\left(\begin{array}{cccc}
			\frac{1}{4} \lambda_{\Phi} & \frac{1}{2} \lambda_{\Phi\chi} & \frac{1}{2}\lambda_{\phi\Phi} & 0 \\
			& \frac{1}{4} \lambda_{\chi} & \frac{1}{2}\lambda_{\phi\chi} & 0 \\
			& & \frac{1}{4}\lambda_{\phi} &\frac{1}{2}\lambda_{\phi H} \\
			& & & \lambda_{H}
		\end{array}\right)\,.
	\end{equation}
	To ensure the potential is bounded from below, the matrix needs to be copositive~\cite{PING1993109,Chakrabortty:2013mha}. The vacuum stability conditions are summarized as
	\begin{equation}		
		\lambda_{\Phi}\geqslant 0\,,\lambda_{\chi}\geqslant0\,,\lambda_{\phi}\geqslant0\,,\lambda_{H}\geqslant0\,,
	\end{equation}
	and	for the case $\lambda_{\phi H}>0$, 	
	\begin{equation}
		\lambda_{\Phi\chi}+\sqrt{\lambda_{\Phi}\lambda_{\chi}/2}\geqslant0\,.
	\end{equation}
	While for the case $\lambda_{\phi H}\leqslant0$, one needs	
	\begin{align}
		&
		\lambda_{H}\lambda_{\phi}-\lambda_{\phi H}^{2}\geqslant0\,,~~\quad\lambda_{H}\lambda_{\Phi\chi}+\sqrt{\lambda_{\Phi}\lambda_{\chi}\lambda_{H}^{2}/2}\geqslant0\,, \notag\\
		& 
		2\lambda_{H}\lambda_{\phi\Phi}+\sqrt{\lambda_{\Phi}\lambda_{\phi}\lambda_{H}^{2}-\lambda_{\Phi}\lambda_{H}\lambda_{\phi H}^{2}}\geqslant0\,,~~\quad\lambda_{H}\lambda_{\phi\chi}+\sqrt{\lambda_{\chi}\lambda_{\phi}\lambda_{H}^{2}-\lambda_{\chi}\lambda_{H}\lambda_{\phi H}^{2}}\geqslant0\,, \notag\\
		& \sqrt{\frac{1}{2}\lambda_{\Phi}\lambda_{\chi}\lambda_{H}^{2}\left(\lambda_{H}\lambda_{\phi}-\lambda_{\phi H}^{2}\right)}+\lambda_{H}\lambda_{\Phi\chi}\sqrt{\left(\lambda_{H}\lambda_{\phi}-\lambda_{\phi H}^{2}\right)}+\lambda_{H}\lambda_{\phi\Phi}\sqrt{2\lambda_{H}\lambda_{\chi}}\notag\\
		&+\lambda_{H}\lambda_{\phi\chi}\sqrt{\lambda_{\Phi}\lambda_{H}}+\sqrt{\frac{1}{2}\left(2\lambda_{H}\lambda_{\Phi\chi}+\sqrt{2\lambda_{\Phi}\lambda_{\chi}\lambda_{H}^{2}}\right)\left(2\lambda_{H}\lambda_{\phi\Phi}+\sqrt{\lambda_{\Phi}\lambda_{H}\left(\lambda_{H}\lambda_{\phi}-\lambda_{\phi H}^{2}\right)}\right)}\notag\\
		&\times\sqrt{\left(2\lambda_{H}\lambda_{\phi\chi}+\sqrt{2\lambda_{H}\lambda_{\chi}\left(\lambda_{H}\lambda_{\phi}-\lambda_{\phi H}^{2}\right)}\right)}\geqslant0\,.
	\end{align}
	\subsection{Minima conditions}
	Consider the scalar potential of the field $\Phi$,
	\begin{equation}
		\begin{aligned}
			V&\supset-\frac{\mu_{\Phi}^2}{2}\operatorname{Tr}\left[\Phi^{\dagger}\Phi\right]+\frac{\lambda_{\Phi}}{4}(\operatorname{Tr}\left[\Phi^{\dagger}\Phi\right])^{2}\\
			&\supset-\frac{\mu_{\Phi}^2}{2}(\phi^{0})^{2}+\frac{\lambda_{\Phi}}{4}(\phi^{0})^{4}\,.
		\end{aligned}
	\end{equation}
	The minima condition $(\partial V/ \partial \phi_{0})|_{\phi_{0}=v_{D}}=0$ yields $\mu_{\Phi}^{2}=\lambda_{\Phi}v_{D}^{2}$.

	\subsection{Higgs phenomenology}
	After electroweak symmetry breaking, the scalar state $h$ mixes with the scalar state $\phi$. We identify the lighter mass state $h_{1}$ as the SM observed $125~\rm{GeV}$ Higgs boson, and  take the heavier mass state $h_{2}$ to be $200~\rm{GeV}$ for illustration. The coupling of the mass eigenstates to SM particles are
	\begin{equation}
		\mathcal{L}_{\text{int}}^{H-SM} \supset\frac{\cos{\alpha}~h_{1}+\sin{\alpha}~h_{2}}{v}(2m_{W}^{2}W_{\mu}^{+}W^{\mu-}+m_{Z}^{2}Z_{\mu}Z^{\mu}+\sum_{f} m_{f}\bar{f}f)\,.
	\end{equation}
	The Higgs self-interaction in terms of the mass eigenstates are
	\begin{align}
		\mathcal{L}_{\text{self-int}}^{H} &\supset -\kappa_{111}h_{1}^{3}-\kappa_{222}h_{2}^{3}-\kappa_{112}h_{1}^{2}h_{2}-\kappa_{122}h_{1}h_{2}^{2}\,,\notag\\
		\kappa_{111}&=-\lambda v \cos^{3}\alpha-\lambda_{\phi H}v\cos{\alpha}\sin^{2}\alpha+\frac{1}{3}\kappa_{\phi}\sin^{3}\alpha+\frac{1}{2}\kappa_{\phi H}\cos^{2}\alpha\sin{\alpha}\,,\notag\\
		\kappa_{222}&=-\lambda v\sin^{3}\alpha-\lambda_{\phi H}v\cos^{2}\alpha\sin{\alpha}-\frac{1}{3}\kappa_{\phi}\cos^{3}\alpha-\frac{1}{2}\kappa_{\phi H}\cos{\alpha}\sin^{2}\alpha\,,\\
		\kappa_{112}&=\kappa_{\phi H}(\cos{\alpha}\sin^{2}\alpha-\frac{1}{2}\cos^{3}\alpha)-3\lambda v\cos^{2}\alpha\sin{\alpha}+\lambda_{\phi H}v (2\cos^{2}\alpha\sin{\alpha}-\sin^{3}\alpha)\notag\\
		&-\kappa_{\phi}\cos{\alpha}\sin^{2}\alpha\,,\notag\\
		\kappa_{122}&=-\kappa_{\phi H}(\cos^{2}\alpha\sin{\alpha}-\frac{1}{2}\sin^{3}\alpha)-3\lambda v\sin^{2}\alpha\cos{\alpha}+\lambda_{\phi H}v(2\sin^{2}\alpha\cos{\alpha}-\cos^{3}\alpha)\notag\\ &+\kappa_{\phi}\cos^{2}\alpha\sin{\alpha}\,.\notag
	\end{align}
	
	If $m_{\rho_1}<m_{h_1}/2$, the Higgs decay into two DM particles through the tree level, the partial width is
	\begin{equation}
		\Gamma_{h_1\rightarrow\rho_{1}\rho_{1}^{*}}=\frac{\kappa_{\rho_{1}\rho_{1}^{*}h_{1}}^{2}}{16\pi m_{h_1}}\sqrt{1-\frac{4m_{\rho_1}^{2}}{m_{h_1}^{2}}}\,.
	\end{equation}
	The branching ratio of the Higgs invisible decay is 
	\begin{equation}
		B_{\text{inv}}=\frac{\Gamma_{\text{BSM}}}{\Gamma_{\text{BSM}}+\cos^{2}\alpha\Gamma_{\text{SM}}}\,,
	\end{equation}
	where $\Gamma_{\text{SM}}=4.2~\text{MeV}$ is the total Higgs decay width in the SM, and $\Gamma_{\text{BSM}}=\Gamma_{h_1\rightarrow\rho_{1}\rho_{1}^{*}}$ in this case. This sets a constraints on the coupling $\kappa_{\rho_{1}\rho_{1}^{*}h_{1}}$.
	
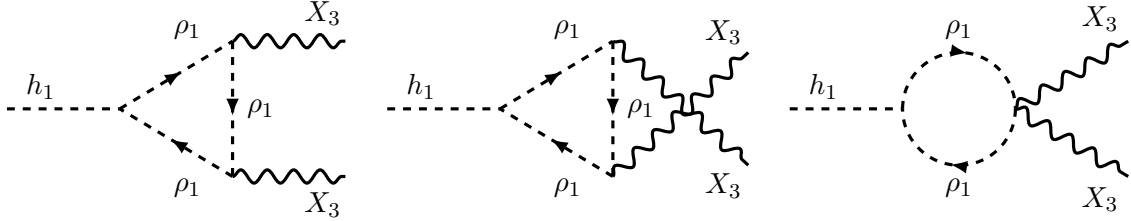
\begin{figure}[!h]%[!t]
		%1   
		\subfigure{
			\begin{minipage}{0.3\textwidth}
				\begin{tikzpicture}[line width=1.2 pt, scale=1.5, >=latex]
					\draw[dashed] (-1,0)--(0,0);	
					\draw[scalar] (0,0)--(1,0.6);
					\draw[scalar] (1,-0.6)--(0,0);
					\draw[scalar] (1,0.6)--(1,-0.6);
					\draw[decorate, decoration={snake}] (1,0.6)--(2,0.6);
					\draw[decorate, decoration={snake}] (1,-0.6)--(2,-0.6);
					\node at (-0.7,0.2) {$h_{1}$};
					\node at (0.6,0.7) {$\rho_{1}$};
					\node at (0.6,-0.7) {$\rho_{1}$};
					\node at (1.25,0) {$\rho_{1}$};
					\node at (1.8,-0.85) {$X_{3}$};
					\node at (1.8,0.85) {$X_{3}$};
					
				\end{tikzpicture}
		\end{minipage}}
		%2   
		\subfigure{
			\begin{minipage}{0.3\textwidth}
				\begin{tikzpicture}[line width=1.2 pt, scale=1.5, >=latex]
					\draw[dashed] (-1,0)--(0,0);	
					\draw[scalar] (0,0)--(1,0.6);
					\draw[scalar] (1,-0.6)--(0,0);
					\draw[scalar] (1,0.6)--(1,-0.6);
					\draw[decorate, decoration={snake}] (1,0.6)--(2.2,-0.5);
					\draw[decorate, decoration={snake}] (1,-0.6)--(2.2,0.5);
					\node at (-0.7,0.2) {$h_{1}$};
					\node at (0.6,0.7) {$\rho_{1}$};
					\node at (0.6,-0.7) {$\rho_{1}$};
					\node at (1.25,0) {$\rho_{1}$};
					\node at (1.98,-0.68) {$X_{3}$};
					\node at (1.98,0.68) {$X_{3}$};
					
				\end{tikzpicture}
		\end{minipage}}
		%3
		\subfigure{
			\begin{minipage}{0.3\textwidth}
				\begin{tikzpicture}[line width=1.2 pt, scale=1.5, >=latex]
					\draw[dashed] (-1,0)--(0,0);	
					\draw[scalar] (0,0) arc (180:0:0.5);
					\draw[scalar] (1,0) arc (0:-180:0.5);
					\draw[decorate, decoration={snake}] (1,0)--(2,0.6);
					\draw[decorate, decoration={snake}] (1,0)--(2,-0.6);
					\node at (-0.7,0.2) {$h_{1}$};
					\node at (0.5,0.7) {$\rho_{1}$};
					\node at (0.5,-0.7) {$\rho_{1}$};
					\node at (1.75,-0.8) {$X_{3}$};
					\node at (1.75,0.8) {$X_{3}$};
				\end{tikzpicture}
		\end{minipage}}
		\caption{Feynman diagrams for the Higgs invisible decay to the dark photons. }
		\label{Higgs invisible decay}
	\end{figure}

	If the DM masses are larger than half of the Higgs boson mass, the invisible decay of the Higgs boson proceeds via $h_{1}\rightarrow X_{3}X_{3}$, through the charge dark scalar loops, as shown in Fig.~\ref{Higgs invisible decay}. We note that the dark scalars running in the loop can also include the heavier scalar $\rho_{2}$. However, due to the mass hierarchy between the two scalars, the contributions from the $\rho_{1}$ loops are dominant. 
	Using the package micrOMEGAs, the calculated Higgs invisible decay width for the benchmark points in Table~\ref{benchmark points} are
	\begin{equation}
		\begin{aligned}
			&\text{BP1:}\qquad\Gamma_{h_1\rightarrow X_{3}X_{3}}=8.0\times 10^{-21}~\rm{MeV}\,,\\
			&\text{BP2:}\qquad\Gamma_{h_1\rightarrow X_{3}X_{3}}=1.5\times 10^{-21}~\rm{MeV}\,,
		\end{aligned}   
	\end{equation}
	which are dominant by the first diagram in the figure and are much smaller than the limit $B_{\text{inv}}<0.26$~\cite{ATLAS:2019cid}.
	
	\subsection{DM direct detection}
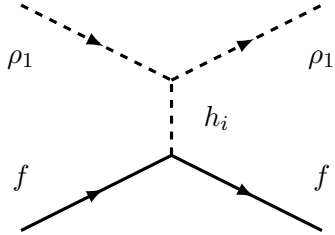
\begin{figure}[!h]%[!t]  
		%1   
		\subfigure{
			\begin{minipage}{0.5\textwidth}
				\begin{tikzpicture}[line width=1.2 pt, scale=2, >=latex]
					\draw[scalar] (-1,0.5)--(0,0);
					\draw[scalar] (0,0)--(1,0.5);
					\draw[dashed] (0,-0.5)--(0,0);
					\draw[fermion] (-1,-1)--(0,-0.5);
					\draw[fermion] (0,-0.5)--(1,-1);
					\node at (-1,0.14) {$\rho_{1}$};
					\node at (-1,-0.65) {$f$};
					\node at (0.3,-0.25) {$h_{i}$};
					\node at (1,0.14) {$\rho_{1}$};
					\node at (1,-0.65) {$f$};
				\end{tikzpicture}
		\end{minipage}}
		\caption{Scattering of the DM with the SM fermions in the $ SU(2)_{D} $ model at the tree level. }
		\label{DM direct detection}
	\end{figure}

	The dominant contribution to the spin-independent contribution of the DM-nucleon elastic scattering cross-section comes from t-channel exchange of the Higgs boson $h_{1}$ and $h_{2}$. The relevant Feynman diagram is shown in Fig.~\ref{DM direct detection}. Since the scattering of $X^{\pm}$ occurs only at the loop level, we focus on the tree-level contribution from the $\rho_{1}^{(*)}$ DM. The effective Lagrangian for the DM interactions with light quarks and gluons can be written as
	\begin{equation}
		\mathcal{L}_{\text{eff}}^{DM-q,g} =
		\sum_{q=u,d,s}C_{q}^{\chi}m_{q}\chi\bar{\chi}q\bar{q}+C_{G}^{\chi}\frac{\alpha_{s}}{\pi}\chi\bar{\chi}G^{a}_{\mu\nu}G^{a\mu\nu}\,,
	\end{equation}
	where $\alpha_{s}$ is the strong coupling constant and the $\chi$ refers to the DM. The coefficient $C_{q}^{\chi}$ of the DM $\rho_{1}^{(*)}$ and the light quark interaction term is
	\begin{equation}
		\begin{aligned}
			C_{q}^{\rho_{1}^{(*)}} &=\frac{m_{q}}{v}\left(\frac{\kappa_{\rho_{1}\rho_{1}^{*}h_{1}}\cos{\alpha}}{m_{h_{1}}^{2}}+\frac{\kappa_{\rho_{1}\rho_{1}^{*}h_{2}}\sin{\alpha}}{m_{h_{2}}^{2}}\right)\,.
		\end{aligned}
	\end{equation}
	The spin-independent cross-section is given by 
	\begin{equation}
		\sigma_{\rho_{1}N}^{\text{SI}}=\frac{\mu_{N}^{2}}{\pi m_{\rho_{1}}^{2}}(F^{\rho_{1}}_{N})^{2}\,,
	\end{equation}
	where ``N" stands for the nucleon and $ \mu_{N}$ is the DM-nucleon reduced mass.  The form factor $F^{\rho_{1}}_{N}$ is related to the scalar couplings~\cite{Shifman:1978zn}
	\begin{equation}
		F^{\rho_{1}}_{N}=\sum_{q=u,d,s}\frac{m_{N}}{m_{q}}C_{q}^{\rho_{1}}f^{N}_{q}+\frac{2}{27}\sum_{q=c,b,t}\frac{m_{N}}{m_{q}}C_{G}^{\rho_{1}}f^{N}_{G}\,.
	\end{equation}
	Here, $f_{q}^{N}\equiv\langle N|m_{q}q\bar{q}|N\rangle$  represents the contribution of quark $q$ to the nucleon mass and $f^{N}_{G}=1-\sum_{q=u,d,s}f^{N}_{q}$. This leads to
	\begin{equation}
		F^{\rho_1}_{N}=\sum_{q=u,d,s}\frac{m_{N}}{v}\left(\frac{\kappa_{\rho_{1}\rho_{1}^{*}h_{1}}\cos{\alpha}}{m_{h_{1}}^{2}}+\frac{\kappa_{\rho_{1}\rho_{1}^{*}h_{2}}\sin{\alpha}}{m_{h_{2}}^{2}}\right)f_{q}^{p}\,.
	\end{equation}
	For numerical calculations, we use $f_{p}^{d}=0.0191$, $f_{p}^{u}=0.0153$, $f_{p}^{s}=0.0447$, $f_{n}^{d}=0.0273$, $f_{n}^{u}=0.011$, $f_{n}^{s}=0.0447$, and $f_{p}^{G}=0.925$ from~\cite{ParticleDataGroup:2012pjm,Belanger:2013oya}. The explicit form of the $\kappa$ parameters are provided in the Appendix.
	
\begin{figure}[!ht]
		\begin{minipage}{0.49\linewidth}
			\centerline{\includegraphics[width=1.3\textwidth]{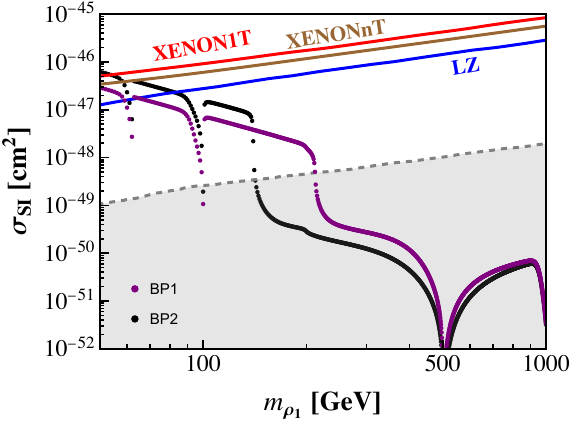}}
		\end{minipage}
		\caption{The spin-independent DM-nucleon scattering cross-section as a function of the DM mass $m_{\rho_1}$ as well as the exclusion limits from some direct detection experiments. In our analysis, the dark sector coupling and other relevant parameters are fixed at the benchmark points listed in Table.~\ref{benchmark points}. Specifically, we choose BP1, corresponding to the purple dots and BP2 corresponding to the black dots. These benchmark choices lead to the mass of the other DM component, $m_{X^{+}}=\sqrt{2}g_{D}v_{D}$ as around $212~\rm{GeV}$ and $140~\rm{GeV}$ respectively. The points shown in the figure are selected by imposing the relic density. The red line is the limit line from XENON1T~\cite{XENON:2018voc}, the brown line is from XENONnT~\cite{XENON:2023cxc}, and the blue line is from LZ~\cite{LZ:2022lsv}. The corresponding direct detection experiments exclude the upper region of the colored line. The gray region at the bottom is the neutrino background~\cite{Ruppin:2014bra}.}
		\label{DD}
	\end{figure}

	 In the multi-component scenario, the predicted event should be rescaled by the fraction of DM relic density by the component under consideration. We approximate the local dark matter density ratio to the ratio of dark matter relic density, $r_{i}=\Omega_{i}h^{2}/(\Omega_{X^{\pm}}h^{2}+\Omega_{\rho_{1}^{(*)}}h^{2})$, $i=X^{\pm},\rho_{1}^{(*)}$. The experimental bound should consider the rescaling as 
	\begin{equation}
		\sum_{i=X^{\pm},\rho_{1}^{(*)}}r_{i}\sigma_{i}^{SI}<\sigma_{\text{limit}}(m_{\rho_{1}})\,.
	\end{equation}
	The spin-independent cross-section for this model, along with the limitations of experiments  XENON1T~\cite{XENON:2018voc}, XENONnT~\cite{XENON:2023cxc} and  LZ~\cite{LZ:2022lsv} are shown in Fig.~\ref{DD}. The cross-section results for our benchmark points are plotted. As the mass of $\rho_1$ increases, its contribution to the total relic density decreases, resulting in an overall downward trend. The two dips in the low-mass region arise from the Higgs-resonance region, $m_{\rho_1}\simeq m_{h_{i}}/2$. The dip around $500~\text{GeV}$ is caused by the resonance region of $\rho_{0}$, where the reduced relic density fraction leads to a suppression, corresponding to $m_{\rho_1}\simeq m_{\rho_0}/2$.
	
	\subsection{DM indirect detection}
	Previous analyses of dark matter induced emission from dwarf spheroidal galaxies (dSphs) with the Fermi Large Area Telescope (LAT) have provided the most stringent and robust constraints on the dark matter annihilation cross-section and mass~\cite{Fermi-LAT:2015att,Fermi-LAT:2016uux,DiMauro:2021qcf,McDaniel:2023bju}.
	
\begin{figure}[!ht]
		\begin{minipage}{0.49\linewidth}
			\centerline{\includegraphics[width=1.3\textwidth]{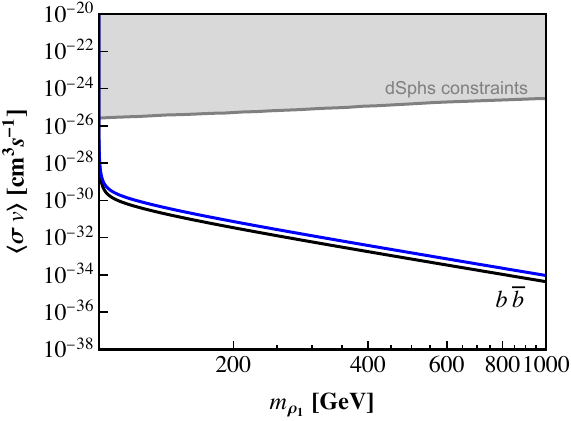}}
		\end{minipage}
		\caption{Indirect detection limit on the DM self-annihilation process $\rho_{1}\rho_{1}^{*}\rightarrow b\bar{b}$. The gray line represents the limit considering the combined analysis of six years of observations of 30 dwarf spheroidal galaxies (dSphs), while the shaded region represents the parameter space excluded.}
		\label{ID}
	\end{figure}

	We utilize the indirect search limits on DM self-annihilation cross section from~\cite{McDaniel:2023bju} which combined analysis of J-factors and uncertainties for 30 dwarf spheroidal galaxies (dSphs) from the Milky Way dwarf spheroidal galaxies with 14 years of Fermi-LAT data. The gray line in Fig.~\ref{ID} represents the exclusion limit from~\cite{McDaniel:2023bju}. The gray-shaded region is excluded by the Fermi-LAT data limit on DM self-annihilation into bottom-quark pairs. The black and blue lines show the cross-section of DM self-annihilation into bottom pairs for the parameters from BP1 and BP2, respectively. Both benchmark points lie well below the current experimental limits. The displayed values are obtained from Eq.(\ref{DM-bbar}) in the Appendix. For the massive dark matter particles considered here, the cross-section into $b\bar{b}$ shows values much below the experimental sensitivity, for the reason that it is significantly suppressed compared with the process into gauge-boson final states once these channels are open.
	
	\subsection{Dark radiation}
	We assume that both the visible and dark sector thermalized to be in equilibrium in the early Universe. Suppose the dark gauge coupling $g_{D}$ overtakes other couplings, then the $X^{\pm}$ annihilation process is dominantly taken over by the two dark photon final states (via the last three diagrams in Fig.~\ref{DM annihilation}), with a cross-section as
	\begin{equation}
		\langle\sigma v \rangle_{X^{+}X^{-}\rightarrow X_{3}X_{3}}=\frac{\pi g_{D}^2}{4m_{X^{\pm}}^{2}}\,.
	\end{equation}
	The produced dark photon contribute to the radiation energy density, which is constrained by both BBN and CMB observations via the total effective number of neutrino species, defined as $N_{\text{eff}} = N^{\text{SM}}_{\text{eff}} + \Delta N_{\text{eff}}$. The energy density from the dark photon can be expressed as
	\begin{equation}
		\rho_{\gamma_{D}}=\frac{\pi^2}{30}g_{\gamma_{D}}T_{D}^{4}\,.
	\end{equation}
	The contribution $ \Delta N_{\text{eff}}$ relative to the density $\rho_{\nu_{i}}$ of a single SM neutrino is~\cite{Wallisch:2018rzj}
	\begin{equation}
		\Delta N_{\text{eff}}\equiv\frac{\rho_{\gamma_{D}}}{\rho_{\nu_{i}}}=\frac{8}{7}\left(\frac{T^{0}}{T_{\nu}^{0}}\right)^4\frac{\rho^{0}_{\gamma_{D}}}{\rho^{0}_{\gamma}}\,,
		\label{DeltaNeff}
	\end{equation}
	where the index ``0" stands for today. Using the observational constraints on $ N_{\text{eff}} $~\cite{Fields:2019pfx} and the SM contribution $ N^{\text{SM}}_{\text{eff}}=3.044$~\cite{Akita:2020szl,Bennett:2020zkv}\footnote{$ N_{\text{eff}} $ differs from the integer three, due to known properties of neutrinos in the early Universe. First, accurate calculations show that neutrinos are still interacting with the primordial plasma when the process of electron-positron annihilation begins which gives a contribution $ \Delta N_{\text{eff}}\simeq 0.03 $~\cite{Giunti:2007ry}. Second, the energy dependence of neutrino interactions causes the high-temperature tail of the Fermi-Dirac distribution to interact more strongly, leading to an energy-dependent distortion in the energy spectrum of neutrino gas which leads to $ \Delta N_{\text{eff}}\simeq 0.01 $~\cite{Giunti:2007ry}. These effects conspire to raise $ N_{\text{eff}} $~\cite{Giunti:2007ry,TopicalConvenersKNAbazajianJECarlstromATLee:2013bxd}.}
	, we find the $2~\sigma$ upper limits: 
	\begin{equation}
		\begin{aligned}
			&N_{\text{eff}}^{\text{CMB}}=2.859\pm0.314\quad	&\rightarrow \quad \quad &\Delta N_{\text{eff}}^{\text{CMB}}<0.443\,,\\
			&N_{\text{eff}}^{\text{BBN}}=2.878 \pm0.278\quad	&\rightarrow \quad \quad &\Delta N_{\text{eff}}^{\text{BBN}}<0.390\,,\\
			&N_{\text{eff}}^{\text{CMB+BBN}}=2.862\pm0.153\quad	&\rightarrow \quad \quad &\Delta N_{\text{eff}}^{\text{CMB+BBN}}<0.124\,.
		\end{aligned}
	\end{equation}
	Assuming the two sector decouple at $ T^{\text{dec}} $, and that the entropy in each sector is conserved separately thereafter. The ratio of temperatures at any time $ t $ and $ t^{\prime} $ after $ T^{\text{dec}}$ is:
	\begin{equation}\label{s conservation}
		\frac{T_{i}^{t}}{T_{i}^{t^{\prime}}}=\left[\frac{g^{*s}_{i}(T_{i}^{t^{\prime}})}{g^{*s}_{i}(T_{i}^{t})}\right]^{1/3}\frac{a(T_{i}^{t^{\prime}})}{a(T_{i}^{t})}\,,
	\end{equation} 
	where $i$ represents for the sector, $i=D~\rm{or}~SM$, and $a$ is the time-dependent scale factor, $T_{D}$ refers to the temperature of the dark sector and $ g^{*s}_{i}$ are the effective number of relativistic degrees of freedom entering into the entropy density. Noting that the scale factor varies only by time. Then we take $t^{\prime}$ as the time after the two sectors decouple completely finished, $T_{D}^{t^{\prime}}=T_{D}$, $T_{\text{SM}}^{t^{\prime}}=T$, via Eq.~(\ref{s conservation}), the ratio of temperature between the dark sector and the SM today is given by
	\begin{equation}
		\frac{T_{D}^{0}}{T^{0}}=\frac{T_{D}}{T}\left[\frac{g^{*s}_{D}(T_{D})}{g^{*s}_{D}(T_{D}^{0})}\right]^{1/3}\left[\frac{g^{*s}_{\text{SM}}(T)}{g^{*s}_{\text{SM}}(T^{0})}\right]^{-1/3}\,.
	\end{equation} 
	Therefore, we take $ t=0 $ and $ t^{\prime}=t_{\text{dec}} $ in Eq.~(\ref{s conservation}), using also $ \rho^{0}_{\gamma_{D}}/\rho^{0}_{\gamma}=(g_{\gamma_{D}}/g_{\gamma})(T_{D}^{0}/T^{0})^4 $, we get that Eq.~(\ref{DeltaNeff}) becomes
	\begin{equation}
		\Delta N_{\text{eff}}=\frac{8}{7}\left(\frac{11}{4}\right)^{4/3}\frac{g_{\gamma_{D}}}{g_{\gamma}}\left[\frac{g^{*s}_{D}(T_{D}^{\text{dec}})}{g^{*s}_{D}(T_{D}^{0})}\right]^{4/3}\left[\frac{g^{*s}_{\text{SM}}(T^{\text{dec}})}{g^{*s}_{\text{SM}}(T^{0})}\right]^{-4/3}\,,
	\end{equation}
	and we consider the dark sector is not reheated after both sector decouple from each other, we have $g_{D}^{*s}(T_{D}^{\text{dec}})=g^{*s}_{D}(T_{D}^{0})$, then we obtain
	\begin{equation}
		\Delta N_{\text{eff}}=\frac{8}{7}\left(\frac{11}{4}\right)^{4/3}\left[\frac{g^{*s}_{\text{SM}}(T^{\text{dec}})}{43/11}\right]^{-4/3}\,,
	\end{equation}
	where $ g_{\gamma_{D}}=g_{\gamma}=2 $, $ g_{\text{SM}}^{*s}(T^0)=43/11 $. Requiring $ \Delta N_{\text{eff}} $ not to exceed 0.124, we get the lower bound of $ T^{\text{dec}} $ as
	\begin{equation}
		g^{*s}_{\text{SM}}(T^{\text{dec}})\gtrsim56.8\quad\rightarrow\quad T^{\text{dec}}\gtrsim375~\rm{MeV}\,.
	\end{equation}
	The decoupling temperature of the two sectors is set by the freeze-out temperature of the light DM species at $T^{\text{dec}}=m_{\chi}/x_{f}$, take $x_{f}\sim 20$, we can get the lower limit of the DM mass $m_{\chi}\gtrsim 7.5~\rm{GeV}$. 
	\subsection{Ellipticity}
	The self-interaction mediated by the long-range force in the hidden sector can impact the form of the non-linear structure. Through cosmological observations of the ellipticity of the galaxy halos, one can get constraints on the DM self-interaction~\cite{Feng:2009mn}.
	
	Elastic scattering between DM in the halos of galaxies transfers energy and shapes the constant-density contours, which form a central region with constant density called the core. In fact, if the collisions are fast enough to create $\mathcal{O}$(1) changes to the energy of the DM particles in the halo, it will heat up the core, eliminate speed correlations and lead to a more and more rounder core. It will also drive the halo to isothermality, which means heat is conducted from the hotter outer to the cooler inner parts of the halo through collisions. In cosmology, we define a timescale as the relaxation time $ \tau_{\rm{iso}} $ on which the velocity vector fully randomizes. Over periods longer than the relaxation time, in collisions, the ejection of DM particles from the parent halo will further cool the core. For an isolated halo, this will eventually bring about core collapse.
	
	We concentrate on the timescale for an average DM particle to change its kinetic energy $ E_k $ by an $\mathcal{O}$(1) factor and interpret it as the relaxation timescale, by which we can use constraints from measurements of the elliptical galaxy halos to put limits on the dark sector interaction. We consider a large initial ellipticity decreases scenario, the explicit calculation of the time to erase ellipticity for different measured ellipticity values and other details can be seen in~\cite{Agrawal:2016quu}.
	
	The timescale is defined as
	\begin{equation}
		\tau_{\rm{r}}=\frac{E_{k}}{\langle\dot{E_{k}}\rangle}\,,
	\end{equation}
	where the $\langle \cdot \rangle$ denotes thermal averaging, and the mean energy change rate of a typical particle is
	\begin{equation}
		\langle\dot{E}_k\rangle=\int \delta E_{k}d\Omega\frac{d\sigma}{d\Omega} v n_{\chi}  f(v) d^3v\,.
		\label{mean energy change}
	\end{equation}
	We assume the DM particles in the halos have a Maxwell-Boltzmann velocity distribution $ f(v) $. With the Rutherford scattering formula
	\begin{equation}
		\frac{d\sigma}{d\Omega}
		=\frac{\alpha_{D}^2}{4m_{\chi}^2v^4\sin^4(\beta/2)}\,,
	\end{equation}
	and the kinetic energy exchange in each collision $ \delta E_k=E_k(1-\cos\beta) $ with $ E_k=m_{\chi}v^2/2 $, the Eq.~(\ref{mean energy change}) gives
	\begin{equation}
		\begin{aligned}
			\langle\dot{E}_k\rangle
			&=\int \delta E_{k}n_{\chi} \frac{d\sigma}{d\Omega} v f(v) d^3v d\Omega\\
			&=\frac{\alpha_{D}^2\rho_{\chi}\sqrt{\pi}}{m_{\chi}^2 v_{0}^{3}}\int \frac{1-\cos\beta}{\sin^4(\beta/2)}v e^{-v^2/v_{0}^2}dv~d\cos\beta\\
			&=-\frac{2\alpha_{D}^2\rho_{\chi}\sqrt{\pi}}{m_{\chi}^2 v_{0}^{3}}\ln(1-\cos{\beta_{\rm{min}}})\,,
		\end{aligned}
	\end{equation}
	where $\beta$ refers to the scattering angle in the lab frame. The minimum scattering angle is related to the maximum impact parameter through
	\begin{equation}
		b_{\rm{max}}=\frac{\alpha_{D}}{m_{\chi}v_{0}^{2}}\cot(\beta_{\rm{min}}/2)\,,
	\end{equation}
	where the $ b_{\rm{max}} $ should be chosen as the Debye screening length. When the scale under discussion is larger than the Debye length, the colliding particles can be viewed as overall electrically neutral, and conversely, as charged one. The Debye length can be expressed as
	\begin{equation}
		\lambda_{D}\sim\frac{m_{\chi}v_{0}}{\sqrt{4\pi\alpha_{D}\rho_{\chi}}}\,.
	\end{equation}
	We get constraints on the $\alpha_{D}$ by requiring the relaxation timescale to be larger than the age of the Universe since the ellipticity of the galaxies considered is non-zero,
	\begin{equation}
		\tau_{\mathrm{r}} \equiv \frac{E_k} {\langle\dot{E}_k\rangle} \simeq \frac{m_{\chi}^3 v_0^3}{4 \sqrt{\pi} \alpha_{D}^2 \rho_{\chi}}\left(\ln \left(\frac{\left(b_{\max } m_\chi v_0^2 \alpha_{D}^{-1}\right)^2+1}{2}\right)\right)^{-1} \geqslant 10^{10} \text { years }\,.
	\end{equation}
	The ``Coulomb logarithm'' is about $90$, and the DM density $ \rho_{\chi}(r) $ is taken at a radius $ \sim(3-10)~\rm{kpc} $, where the density drops from 3.5 to 0.7 $ \rm{GeV/cm}^3 $. The velocity varies from 270 to 250 $ \rm{km/s} $. Consider the data from NGC720~\cite{Buote:2002wd}, we get the constraints of $ \alpha_{D}=g_{D}^{2}/4\pi $ as
	\begin{equation}
		\alpha_{D}\leq0.8\sqrt{10^{-11}~(m_{\chi}/\rm{GeV})^3}\,.
	\end{equation}
	
	\section{Conclusions}
	In this work, we extend the SM with an $SU(2)_{D}$ dark sector containing a triplet scalar, a doublet scalar and a singlet scalar. The dark sector couples to the SM via a Higgs portal, which arises from the mixing between the SM Higgs boson and the dark singlet. The dark sector particles get a residual $Z_{3}$ symmetry after the spontaneous symmetry breaking of dark gauge symmetry. The residual $Z_{3}$ symmetry assures the stability of the lightest dark charged particles as the DM. We identify the dark gauge boson and the lighter mass eigenstate of the dark doublet components as DM and work out the existing constraints on our model parameters from unitarity, minima conditions, vacuum stability, Higgs phenomenology, dark radiation, the ellipticity, and the relic density. To demonstrate, we provide two specific benchmark points of the model that satisfy all the constraints considered above. The model offers rich phenomenology, and our analysis of direct and indirect detection prospects shows that significant parameter space remains viable and detectable in future experiments. 
	
	\noindent{\bf Acknowledgements.}
	W.-w. Jiang thank Ze-Kun Liu for useful discussions. This work is supported in part by the National Science Foundation of China (12575105, 12175082).
	
	\appendix 
	\section{Thermal averages of the related interactions in the model}
	In this section, We present a full set of exact, analytic expressions for the total thermal average cross-section of the related $2\rightarrow2$ processes in the model.
	
	In the calculation, we assume that kinetic equilibrium is obtained during freeze-out. According to the definition of particle number density, we have
	\begin{equation}
		f_{i}(\textbf{p}_{i})=\frac{n_{i}}{n_{i}^{\rm{eq}}}f_{i}^{\rm{eq}}(\textbf{p}_{i})\,,
	\end{equation}
	here $f^{\rm{eq}}_{i}(\textbf{p}_i)$ stands for the thermal distribution for the particles $i$ when it is in the thermal equilibrium. The thermal averaged cross-section is defined as 
	\begin{equation}
		\begin{aligned}
			\langle\sigma v_{\text{rel}}\rangle_{ab\rightarrow cd}
			&\equiv\frac{\int \sigma v_{\text{rel}} f^{\rm{eq}}_{a}(\textbf{p}_a)f^{\rm{eq}}_{b}(\textbf{p}_b)d^{3}p_{a}d^{3}p_{b}}{\int f^{\rm{eq}}_{a}(\textbf{p}_a)f^{\rm{eq}}_{b}(\textbf{p}_b)d^{3}p_{a}d^{3}p_{b}}\\
			&=\frac{Tg_{a}g_{b}}{32\pi^{4} n_{a}^{\rm{eq}}n_{b}^{\rm{eq}}}\int_{\tilde{s}_{min}}^{\infty} \frac{\lambda(\sqrt{\tilde{s}},m_{a},m_{b})}{\tilde{s}^{1/2}} K_{1}\lbrack \sqrt{\tilde{s}}/T\rbrack \sum_{a,b,c,d}C_{ab}\sigma_{ab\rightarrow cd}(\tilde{s})d\tilde{s}\,,
			\label{thermal cs}
		\end{aligned} 
	\end{equation}
	here $\tilde{s}=-(\textbf{p}_a+\textbf{p}_a)^{2}=m_{a}^{2}+m_{b}^{2}+2E_{a}E_{b}-2p_{a}p_{b}\cos{\theta_{ab}}$ is the Mandelstam variable. With the on shell condition, the relative velocity between the initial particles $a$ and $b$ is 
	\begin{equation}
		v_{\text{rel}}\equiv\frac{\sqrt{(\textbf{p}_a\cdot \textbf{p}_b)^{2}-m_{a}^{2}m_{b}^{2}}}{E_{a}E_{b}}\xrightarrow{\mathbf{p}_i^2 = m_i^2}\frac{\lambda^{1/2}(\sqrt{\tilde{s}},m_{a},m_{b})}{2E_{a}E_{b}}\,,
	\end{equation}
	and the function $\lambda(x,y,z)$ is defined as
	\begin{equation}
		\lambda(x,y,z)=\lbrack x^{2}-(y+z)^{2}\rbrack\lbrack x^{2}-(y-z)^{2}\rbrack\,,
	\end{equation}
	and $g_{i}$ represents the number of degrees of freedom of particles. $C_{ab}$ is a symmetry factor, $C_{ab}=1/2$ if $a=b$, otherwise $C_{ab}=1$. $K_{1}\lbrack \sqrt{\tilde{s}}/T\rbrack$ is the modified Bessel functions of the second kind and $p_{a,b}$ is the momentum of the incoming particles $a$ and $b$ in their center-of-mass frame. We consider all particles which are in thermal equilibrium follow Maxwell-Boltzmann distributions, $f^{\rm{eq}}_{i}\propto \exp(-E_{i}/T)$. The equilibrium number density is then given by
	\begin{equation}
		n_{i}^{\rm{eq}}=g_{i}\int\frac{d^3p}{(2\pi)^3}f^{\rm{eq}}_{i}(\textbf{p}_{i})=\frac{g_{i}}{2\pi^{2}}m_{i}^{2}TK_{2}\lbrack m_{i}/T\rbrack\,.
	\end{equation}
	The cross section in Eq.~(\ref{thermal cs}) is a function of $\tilde{s}$ in the center of mass system written by
	\begin{equation}
		\sigma_{ab\rightarrow cd}(\tilde{s})=\frac{	\left|\mathcal{M}_{ab\rightarrow cd}\right|^2}{64\pi^{2}\tilde{s}}\frac{p_{c,cm}}{p_{a,cm}}=\frac{\lambda^{1/2}(\sqrt{\tilde{s}},m_c,m_d)}{\lambda^{1/2}(\sqrt{\tilde{s}},m_a,m_b)}\frac{	\left|\mathcal{M}_{ab\rightarrow cd}\right|^2}{g_{a}g_{b}16\pi^{2}\tilde{s}}\,.
	\end{equation}
	We derive the expression for the thermal average cross-section in Eq.~(\ref{thermal cs}) in the s-wave limit. For numerical results, we precisely calculate the relic density of the DM using the package micrOMEGAs. 
	
	The corresponding cross-section for annihilation to the SM fermions is
	\begin{equation}
		\sigma v_{\text{rel}}(\rho_{1}\rho_{1}^{*}\rightarrow f\bar{f})=\frac{N_{c}m_{f}^{2}}{8\pi v^{2}}\left(1-\frac{4m_{f}^{2}}{\tilde{s}}\right)^{\frac{3}{2}}\times\left[\frac{\cos{\alpha}~\kappa_{\rho_{1}\rho_{1}^{*}h_{1}}}{\tilde{s}-m_{h_{1}}^{2}}+\frac{\sin{\alpha}~\kappa_{\rho_{1}\rho_{1}^{*}h_{2}}}{\tilde{s}-m_{h_{2}}^{2}}\right]^2\,,\label{DM-bbar}
	\end{equation}
	with $\tilde{s}=4m_{\rho_{1}}^{2}$ and
	\begin{equation}
		\begin{aligned}
			&
			\kappa_{\rho_{1}\rho_{1}^{*}h_{1}}=\kappa_{\phi\chi}\cos{2\theta}\sin{\alpha}-\sqrt{2}\lambda_{s1}v_{D}\sin{2\theta}\sin{\alpha}-\frac{1}{\sqrt{2}}\lambda_{s2}v_{D}\cos{2\theta}\sin{\alpha}\,,\\
			&
			\kappa_{\rho_{1}\rho_{1}^{*}h_{2}}=-\kappa_{\phi\chi}\cos{2\theta}\cos{\alpha}+\sqrt{2}\lambda_{s1}v_{D}\sin{2\theta}\cos{\alpha}+\frac{1}{\sqrt{2}}\lambda_{s2}v_{D}\cos{2\theta}\cos{\alpha}\,.  
		\end{aligned}
		\label{kappa}
	\end{equation}
	The DM annihilation cross-section into the gauge bosons $W^{+}W^{-}$ and $ZZ$ is
	\begin{equation}
		\begin{aligned}
			\sigma v_{\text{rel}}&(\rho_{1}\rho_{1}^{*}\rightarrow VV)=
			\frac{1}{8\pi v^{2}\tilde{s} }\left[\frac{\cos{\alpha}~\kappa_{\rho_{1}\rho_{1}^{*}h_{1}}}{\tilde{s}-m_{h_{1}}^{2}}+\frac{\sin{\alpha}~\kappa_{\rho_{1}\rho_{1}^{*}h_{2}}}{\tilde{s}-m_{h_{2}}^{2}}\right]^{2}\\
			&
			\times S_{V}\left((\tilde{s}-2m_{V}^2)^{2}+8m_{V}^{4}\right)\left(1-\frac{4m_{V}^{2}}{\tilde{s}}\right)^{\frac{1}{2}} \,,
		\end{aligned}
	\end{equation}
	with $\tilde{s}=4m_{\rho_{1}}^{2}$ and $S_{W}=1$, $S_{Z}=1/2$ are the symmetry factors of final state particles. The DM annihilation cross -section into the Higgs bosons can be written as
	\begin{align}
		\sigma v_{\text{rel}}&(\rho_{1}\rho_{1}^{*}\rightarrow h_{1}h_{1})=\frac{1}{8\pi^{2}\tilde{s}}\left(1-\frac{4m_{h_{1}}^{2}}{\tilde{s}}\right)^{\frac{1}{2}}
		\left[4\kappa_{\rho_{1}\rho_{1}^{*} h_{1} h_{1}}^{2}+\frac{4\kappa_{\rho_{1}\rho_{1}^{*}\rho_{0}}^{2}\kappa_{\rho_{0} h_{1} h_{1}}^{2}}{(\tilde{s}-m_{\rho_{0}}^{2})^{2}}+\frac{36\kappa_{\rho_{1}\rho_{1}^{*} h_{1}}^{2}\kappa_{111}^{2}}{(\tilde{s}-m_{h_{1}}^{2})^{2}}\right.\notag\\
		&
		+\frac{4\kappa_{\rho_{1}\rho_{1}^{*} h_{2}}^{2}\kappa_{112}^{2}}{(\tilde{s}-m_{h_{2}}^{2})^{2}}+\frac{\kappa_{\rho_{1}\rho_{1}^{*} h_{1}}^{4}}{(t-m_{\rho_{1}}^{2})^{2}}+\frac{\kappa_{\rho_{1}\rho_{2}^{*} h_{1}}^{4}}{(t-m_{\rho_{2}}^{2})^{2}}+\frac{\kappa_{\rho_{1}\rho_{1}^{*} h_{1}}^{4}}{(u-m_{\rho_{1}}^{2})^{2}}+\frac{\kappa_{\rho_{1}\rho_{2}^{*} h_{1}}^{4}}{(u-m_{\rho_{2}}^{2})^{2}}\notag\\
		&+\frac{8\kappa_{\rho_{1}\rho_{1}^{*}\rho}\kappa_{\rho_{0} h_{1} h_{1}}\kappa_{\rho_{1}\rho_{1}^{*} h_{1} h_{1}}}{\tilde{s}-m_{\rho_{0}}^{2}}+\frac{12\kappa_{\rho_{1}\rho_{1}^{*} h_{1}}\kappa_{111}\kappa_{\rho_{1}\rho_{1}^{*} h_{1} h_{1}}}{\tilde{s}-m_{h_{1}}^{2}}+\frac{4\kappa_{\rho_{1}\rho_{1}^{*} h_{2}}\kappa_{112}\kappa_{\rho_{1}\rho_{1}^{*} h_{1} h_{1}}}{\tilde{s}-m_{h_{2}}^{2}}\notag\\
		&+\frac{\kappa_{\rho_{1}\rho_{1}^{*} h_{1}}^{2}\kappa_{\rho_{1}\rho_{1}^{*} h_{1} h_{1}}}{t-m_{\rho_{1}}^{2}}+\frac{\kappa_{\rho_{1}\rho_{2}^{*} h_{1}}^{2}\kappa_{\rho_{1}\rho_{1}^{*} h_{1} h_{1}}}{t-m_{\rho_{2}}^{2}}+\frac{\kappa_{\rho_{1}\rho_{1}^{*} h_{1}}^{2}\kappa_{\rho_{1}\rho_{1}^{*} h_{1} h_{1}}}{u-m_{\rho_{1}}^{2}}+\frac{\kappa_{\rho_{1}\rho_{2}^{*}h_{1}}^{2}\kappa_{\rho_{1}\rho_{1}^{*} h_{1} h_{1}}}{u-m_{\rho_{2}}^{2}}\notag\\
		&+\frac{24\kappa_{\rho_{1}\rho_{1}^{*}\rho_{0}}\kappa_{\rho_{0} h_{1} h_{1}}\kappa_{\rho_{1}\rho_{1}^{*} h_{1}}\kappa_{111}}{(\tilde{s}-m_{\rho_{0}}^{2})(\tilde{s}-m_{h_{1}}^{2})}+\frac{8\kappa_{\rho_{1}\rho_{1}^{*}\rho_{0}}\kappa_{\rho_{0} h_{1} h_{1}}\kappa_{\rho_{1}\rho_{1}^{*} h_{1}}\kappa_{112}}{(\tilde{s}-m_{\rho_{0}}^{2})(\tilde{s}-m_{h_{2}}^{2})}+\frac{4\kappa_{\rho_{1}\rho_{1}^{*}\rho_{0}}\kappa_{\rho_{0} h_{1} h_{1}}\kappa_{\rho_{1}\rho_{1}^{*} h_{1}}^{2}}{(\tilde{s}-m_{\rho_{0}}^{2})(t-m_{\rho_{1}}^{2})}\notag\\
		&+\frac{4\kappa_{\rho_{1}\rho_{1}^{*} \rho_{0}}\kappa_{\rho_{0} h_{1} h_{1}}\kappa_{\rho_{1}\rho_{2}^{*} h_{1}}^{2}}{(\tilde{s}-m_{\rho_{0}}^{2})(t-m_{\rho_{2}}^{2})}+\frac{4\kappa_{\rho_{1}\rho_{1}^{*} \rho_{0}}\kappa_{\rho_{0} h_{1} h_{1}}\kappa_{\rho_{1}\rho_{1}^{*} h_{1}}^{2}}{(\tilde{s}-m_{\rho_{0}}^{2})(u-m_{\rho_{1}}^{2})}+\frac{4\kappa_{\rho_{1}\rho_{1}^{*} \rho_{0}}\kappa_{\rho_{0} h_{1} h_{1}}\kappa_{\rho_{1}\rho_{2}^{*}h_{1}}^{2}}{(\tilde{s}-m_{\rho_{0}}^{2})(u-m_{\rho_{2}}^{2})}\\
		&+\frac{24\kappa_{\rho_{1}\rho_{1}^{*} h_{1}}\kappa_{111}\kappa_{\rho_{1}\rho_{1}^{*} h_{2}}\kappa_{112}}{(\tilde{s}-m_{h_{1}}^{2})(\tilde{s}-m_{h_{2}}^{2})}+\frac{12\kappa_{\rho_{1}\rho_{1}^{*} h_{1}}^{3}\kappa_{111}}{(\tilde{s}-m_{h_{1}}^{2})(t-m_{\rho_{1}}^{2})}+\frac{12\kappa_{\rho_{1}\rho_{1}^{*} h_{1}}\kappa_{111}\kappa_{\rho_{1}\rho_{2}^{*} h_{1}}^{2}}{(\tilde{s}-m_{h_{1}}^{2})(t-m_{\rho_{2}}^{2})}\notag\\
		&+\frac{12\kappa_{\rho_{1}\rho_{1}^{*} h_{1}}^{3}\kappa_{111}}{(\tilde{s}-m_{h_{1}}^{2})(u-m_{\rho_{1}}^{2})}+\frac{12\kappa_{\rho_{1}\rho_{1}^{*} h_{1}}\kappa_{111}\kappa_{\rho_{1}\rho_{2}^{*} h_{1}}^{2}}{(\tilde{s}-m_{h_{1}}^{2})(u-m_{\rho_{2}}^{2})}+\frac{4\kappa_{\rho_{1}\rho_{1}^{*} h_{2}}\kappa_{112}\kappa_{\rho_{1}\rho_{1}^{*} h_{1}}^{2}}{(\tilde{s}-m_{h_{2}}^{2})(t-m_{\rho_{1}}^{2})}+\frac{4\kappa_{\rho_{1}\rho_{1}^{*} h_{2}}\kappa_{112}\kappa_{\rho_{1}\rho_{1}^{*} h_{2}}^{2}}{(\tilde{s}-m_{h_{2}}^{2})(t-m_{\rho_{2}}^{2})}\notag\\
		&+\frac{4\kappa_{\rho_{1}\rho_{1}^{*} h_{2}}\kappa_{112}\kappa_{\rho_{1}\rho_{1}^{*} h_{1}}^{2}}{(\tilde{s}-m_{h_{2}}^{2})(u-m_{h_{1}}^{2})}+\frac{4\kappa_{\rho_{1}\rho_{1}^{*} h_{2}}\kappa_{112}\kappa_{\rho_{1}\rho_{1}^{*} h_{2}}^{2}}{(\tilde{s}-m_{h_{2}}^{2})(u-m_{\rho_{2}}^{2})}+\frac{2\kappa_{\rho_{1}\rho_{1}^{*} h_{1}}^{2}\kappa_{\rho_{1}\rho_{2}^{*} h_{1}}^{2}}{(t-m_{\rho_{1}}^{2})(t-m_{\rho_{1}}^{2})}+\frac{2\kappa_{\rho_{1}\rho_{1}^{*} h_{1}}^{4}}{(t-m_{\rho_{1}}^{2})(u-m_{\rho_{1}}^{2})}\notag\\
		&\left.+\frac{2\kappa_{\rho_{1}\rho_{1}^{*} h_{1}}^{2}\kappa_{\rho_{1}\rho_{2}^{*} h_{1}}^{2}}{(t-m_{\rho_{1}}^{2})(u-m_{\rho_{2}}^{2})}+\frac{2\kappa_{\rho_{1}\rho_{2}^{*} h_{1}}^{2}\kappa_{\rho_{1}\rho_{1}^{*} h_{1}}^{2}}{(t-m_{\rho_{2}}^{2})(u-m_{\rho_{1}}^{2})}+\frac{2\kappa_{\rho_{1}\rho_{2}^{*} h_{1}}^{4}}{(t-m_{\rho_{2}}^{2})(u-m_{\rho_{2}}^{2})}+\frac{2\kappa_{\rho_{1}\rho_{1}^{*} h_{1}}^{2}\kappa_{\rho_{1}\rho_{2}^{*} h_{1}}^{2}}{(u-m_{\rho_{1}}^{2})(u-m_{\rho_{2}}^{2})}\right]\,,\notag
	\end{align}
	with $\tilde{s}=4m_{\rho_{1}}^{2}$, $t,u=m_{W}^{2}-m_{\rho_{1}}^{2}-m_{i}^{2}$, here $m_{i}$ is the mass of the mediator particle. The cross sections for $h_{2}h_{2}$ and $h_{1}h_{2}$ channels can be obtained similarly. The related couplings are
	given as
	\begin{align}
		&
		\kappa_{\rho_{1}\rho_{1}^{*} h_{1} h_{1}}=-\lambda_{\phi\chi}\sin^{2}\alpha\,,\qquad\qquad\qquad
		\kappa_{\rho_{1}\rho_{1}^{*} h_{2} h_{2}}=-\lambda_{\phi\chi}\cos^{2}\alpha\,,\notag\\
		&
		\kappa_{\rho_{1}\rho_{1}^{*} h_{1} h_{2}}=\lambda_{\phi\chi}\sin{2\alpha}\,,
		\qquad\qquad\qquad\quad
		\kappa_{\rho_{0} h_{1} h_{1}}=-2\lambda_{\phi\Phi}v_{D}\sin^{2}\alpha\,,\notag\\
		&
		\kappa_{\rho_{0} h_{2} h_{2}}=-2\lambda_{\phi\Phi}v_{D}\cos^{2}\alpha\,,
		\qquad\qquad~\quad
		\kappa_{\rho_{0} h_{1} h_{2}}=2\lambda_{\phi\Phi}v_{D}\sin{2\alpha}\,,\notag\\ 
		&
		\kappa_{\rho_{1}\rho_{1}^{*}\rho_{0}}=\sqrt{2}\kappa_{1}\sin{2\theta}+\frac{\kappa_{2}}{\sqrt{2}}-2\lambda_{\Phi\chi}v_{D}\,,\notag\\
		&
		\kappa_{\rho_{1}\rho_{2}^{*}h_{1}}=\sqrt{2}\lambda_{s1}v_{D}\cos{2\theta}\sin{\alpha}-\frac{1}{\sqrt{2}}\lambda_{s2}v_{D}\sin{2\theta}\sin{\alpha}\,,\notag\\
		%\displaybreak
		& \kappa_{\rho_{1}\rho_{2}^{*}h_{2}}=-\sqrt{2}\lambda_{s1}v_{D}\cos{2\theta}\cos{\alpha}+\frac{1}{\sqrt{2}}\lambda_{s2}v_{D}\sin{2\theta}\cos{\alpha}\,. \notag
	\end{align}
	For DM candidate $X^{\pm}$, one takes $\tilde{s}=4m_{X^{\pm}}^{2}$ and the annihilation cross-section to Higgs is 
	\begin{equation}
		\sigma v_{\text{rel}}(X^{+}X^{-}\rightarrow  h_{i}h_{i})=\frac{2 v_{D}^{2}g_{D}^{4}\kappa_{\rho_{0} h_{i}h_{i}}^{2}}{(\tilde{s}-m_{\rho_{0}}^{2})^{2}}\frac{1}{\pi \tilde{s}}\left(1-\frac{4m_{h_{i}}^{2}}{\tilde{s}}\right)^{\frac{1}{2}}\left(8+\frac{(\tilde{s}-2m_{X^{\pm}}^{2})^{2}}{m_{X^{\pm}}^{4}}\right)\,,
	\end{equation}
	where $i=1,2$ and
	\begin{equation}
		\sigma v_{\text{rel}}(X^{+}X^{-}\rightarrow  h_{1}h_{2})=\frac{ v_{D}^{2}g_{D}^{4}\kappa_{\rho_{0} h_{1}h_{2}}^{2}}{(\tilde{s}-m_{\rho_{0}}^{2})^{2}}\frac{\lambda^{1/2}(\sqrt{\tilde{s}},m_{h_{1}},m_{h_{2}})}{2\pi \tilde{s}^{2}}\left(8+\frac{(\tilde{s}-2m_{X^{\pm}}^{2})^{2}}{m_{X^{\pm}}^{4}}\right)\,.
	\end{equation}

	\bibliography{references1}
\end{document}